\documentclass[11pt,a4paper]{article}

\pdfoutput=1

\usepackage{jheppub}
\usepackage[T1]{fontenc}
\usepackage{epsfig}
\usepackage{amscd}
\usepackage{verbatim}
\usepackage{latexsym}
\usepackage{amsfonts,amsthm,amsmath,mathtools,wasysym}
\usepackage{amssymb,stmaryrd,textcomp,enumerate,bbm}
\usepackage{color}
\usepackage{xcolor}
\usepackage{booktabs}
\usepackage{float}
\usepackage{graphicx}
\usepackage{lscape}
\usepackage{slashed}
\usepackage{subfigure}
\usepackage{multirow,makecell}
\usepackage{amssymb}
\usepackage{soul}

\numberwithin{equation}{section}

\def \be {\begin{equation}}
\def \ee {\end{equation}}
\def \ba {\begin{array}}
\def \ea {\end{array}}
\def \bea{\begin{eqnarray}}
\def \eea{\end{eqnarray}}
\def \nn {\nonumber}

\def \a {\alpha}
\def \b {\beta}
\def \g {\gamma}

\def \d {\delta}

\def \e {\epsilon}
\def \ve {\varepsilon}
\def \m {\mu}
\def \n {\nu}
\def \k {\kappa}
\def \l {\lambda}
\def \L {\Lambda}
\def \s {\sigma}

\def \r {\rho}

\def \th {\theta}
\def \vth {\vartheta}

\def \t {\tau}
\def \z {\zeta}

\def \mA {\mathcal A}
\def \mB {\mathcal B}

\def \mL {\mathcal L}

\def \mN {\mathcal N}

\def \mP {\mathcal P}
\def \mQ {\mathcal Q}

\def \p {\partial}
\def \f {\frac}
\def \df {\dfrac}

\def \sp {{\slashed\propto}}

\def \lt {\left}
\def \rt {\right}

\def \sr {\sqrt}
\def \td {\tilde}

\def \pp {\propto}

\def \lag {\langle}
\def \rag {\rangle}

\def \hi  {{\hat\imath}}
\def \hj  {{\hat\jmath}}
\def \hk  {{\hat k}}
\def \hl  {{\hat l}}

\def \ho  {{\hat 1}}
\def \hw  {{\hat 2}}

\def \ep {\mathrm{e}}
\def \ii {\mathrm{i}}

\def \Tr {{\textrm{Tr}}}

\def \diag {{\textrm{diag}}}

\def \AdS {{\textrm{AdS}}}

\def \gabjm {{\textrm{AdS}_4\times\textrm{S}^7/\mathbb{Z}_k}}
\def \goabjm {{\textrm{AdS}_4\times\textrm{S}^7/(\mathbb{Z}_{rk}\times\mathbb{Z}_{r})}}

\def \Z {{\textrm{Z}}}

\def \bos {{\textrm{bos}}}
\def \fer {{\textrm{fer}}}

\def \I {{\textrm{I}}}
\def \II {{\textrm{II}}}
\def \III {{\textrm{III}}}
\def \IV {{\textrm{IV}}}

\begin{document}

\title{New BPS Wilson loops in $\mathcal N \textbf{=4}$ circular quiver Chern--Simons--matter theories}
\author{Andrea Mauri}
\author{\!\!,~Silvia Penati}
\author{\!\!,~Jia-ju Zhang}
\affiliation{Dipartimento di Fisica, Universit\`a degli Studi di Milano-Bicocca and INFN, Sezione di Milano-Bicocca, Piazza della Scienza 3, I-20126 Milano, Italy}
\emailAdd{andrea.mauri@mi.infn.it, silvia.penati@mib.infn.it, jiaju.zhang@mib.infn.it}

\abstract{We construct new families of 1/4 BPS Wilson loops in circular quiver $\mathcal N=4$ superconformal Chern--Simons--matter (SCSM) theories in three dimensions. They are defined as the holonomy of superconnections that contain non--trivial couplings to scalar and fermions, and cannot be reduced to block--diagonal matrices. Consequently, the new operators cannot be written in terms of double--node Wilson loops, as the ones considered so far in the literature. For particular values of the couplings the superconnection becomes block--diagonal and we recover the known fermionic 1/4 and 1/2 BPS Wilson loops. The new operators are cohomologically equivalent to bosonic 1/4 BPS Wilson loops and are then amenable of exact evaluation via localization techniques. Moreover, in the case of orbifold ABJM theory we identify the corresponding gravity duals for some of the 1/4 and 1/2 BPS Wilson loops.}


\maketitle

\section{Introduction}

BPS Wilson loops (WLs) in  3D supersymmetric Chern--Simons--matter (SCSM) theories exhibit a rich spectrum of peculiar properties that need to be deeper understood.

Due to dimensional reasons not only scalar but also fermion matter together with ordinary gauge connections can be used to construct WLs as the holonomy of generalized (super)connections \cite{Gaiotto:2007qi,Drukker:2009hy}. This allows to define a large web of BPS operators with different degrees of preserved supersymmetry (SUSY).

Even if protected from UV divergent corrections, BPS WLs can still feature non-trivial vacuum expectation values that are often computed exactly by using localization techniques \cite{Pestun:2007rz,Kapustin:2009kz}. Localization predictions can then be directly checked at weak coupling in perturbation theory. Moreover, in theories that allow for string theory or M--theory dual descriptions, BPS WLs in fundamental representation were shown to be dual to fundamental string or M2--brane configurations \cite{Maldacena:1998im,Rey:1998ik}. Therefore, their expectation values at strong coupling can be computed using the holographic description and the matching with localization results expanded at strong coupling provides a crucial test of the AdS/CFT correspondence.

Given a 3D SCSM theory it is therefore crucial to construct and classify the whole spectrum of BPS WLs, identify their gravity duals and compute their expectation values in diverse regimes. In this paper we aim at giving a complete classification of BPS WLs in  circular quiver $\mN=4$ SCSM theories with alternating Chern-Simons levels. Our classification generalizes the one of \cite{Ouyang:2015bmy} and we are also able to identify the precise gravity dual  configurations for a given sublcass of operators.

The present understanding on BPS WLs in SCSM theories can be summarized as follows. BPS WLs operators in SCSM were first introduced in the seminal work \cite{Gaiotto:2007qi}. There, 1/2 BPS operators for $\mN=2$ SCSM theories and 1/3 BPS WLs in $\mN=3$ SCSM theories were defined   as the holonomy of generalized connections that include couplings to the scalar fields of the theories.

Later on, the construction was generalized to models with a higher degree of supersymmetry. In $\mN=6$ ABJ(M) theories \cite{Aharony:2008ug,Hosomichi:2008jb,Aharony:2008gk} with $U(N)_k \times U(M)_{-k}$ gauge group, dual to M--theory in $\gabjm$ background, the whole classification of BPS WLs was carried out in \cite{Ouyang:2015iza,Ouyang:2015bmy} for timelike infinite straight lines in Minkowski spacetime  and for maximal circles in Euclidean space. The most general WL is 1/6 BPS and corresponds to the holonomy of a superconnection that includes parametric couplings to both scalars and fermions. For particular values of the parameters it reduces to the bosonic 1/6 BPS WL with couplings only to scalar fields \cite{Drukker:2008zx,Chen:2008bp,Rey:2008bh}, which is supposed to be dual to smeared M2--branes \cite{Drukker:2008zx}, and to the fermionic 1/2 BPS WLs, dual to M2--/anti--M2--branes \cite{Drukker:2009hy,Lietti:2017gtc}. The generalization to fermionic BPS WLs with a reduced number of preserved supercharges was considered in \cite{Griguolo:2012iq,Cardinali:2012ru,Kim:2013oza,Bianchi:2014laa,Correa:2014aga}. These are featured by non--trivial {\em latitude} angles in the internal R--symmetry space plus possibly latitude deformations of the Euclidean closed contour.

$\mN=4$ circular quiver SCSM theories with gauge group and levels $\prod_{\ell=1}^r[U(N_{2\ell-1})_k\times U(N_{2\ell})_{-k}]$ were introduced in \cite{Gaiotto:2008sd,Hosomichi:2008jd}.
In the general case of different group ranks they can be obtained via decomposition of the $U(N)_k \times U(M)_{-k}$ ABJ theory with $N = \sum_{\ell=1}^r N_{2\ell -1}$ and $M = \sum_{\ell=1}^{r} N_{2\ell}$ \cite{Benna:2008zy,Lietti:2017gtc}. In the special case of equal ranks one obtains the $\mN=4$ orbifold ABJM theory with $[U(N)_k\times U(N)_{-k}]^r$ gauge group \cite{Benna:2008zy}. This theory has a dual description in terms of M--theory in $\goabjm$ background \cite{Benna:2008zy,Imamura:2008nn,Terashima:2008ba}.

For a class of $\mN=4$ SCSM theories corresponding to circular quivers with alternating levels\footnote{The $\mN=4$ SCSM theories with vanishing levels \cite{Imamura:2008dt,Cooke:2015ila} or corresponding to linear quivers \cite{Gaiotto:2008sd} are not included in the discussion.} 1/4 BPS WLs were constructed in \cite{Ouyang:2015bmy}, which are the holonomy of superconnections that include parametric couplings to scalars and fermions. This was carried out under the assumption that the superconnections can be always written as $2 \times 2$ block diagonal matrices. As a consequence the corresponding WLs, when traced, can be expressed as linear combinations of $r$ double--node operators $W^{(\ell)}$ that are nothing but the holonomy of the  $2 \times 2$ block matrices. Therefore, their study reduces to the study of the generic  $W^{(\ell)}$ confined at two adjacent nodes. For a particular choice of the parameters they give rise to the bosonic 1/4 BPS WL that includes couplings only to scalars and the fermionic 1/2 BPS WL with couplings also to fermions \cite{Ouyang:2015qma,Cooke:2015ila}.

In this paper we provide a more general construction of BPS WLs in $\mN=4$ SCSM theories using two different strategies. To begin with, we consider $[U(N)_k\times U(N)_{-k}]^r$  ABJM orbifold models and construct BPS WLs by performing a direct decomposition of BPS WLs classified in ABJM theory. Although this does not provide a full classification of WLs in orbifold ABJM theory, it allows to obtain a class of BPS operators that is however larger than the one known so far. Moreover, this method has the virtue to automatically provide the M-theory dual description of these new operators as given by M2--/anti--M2--branes wrapping some particular circles in the internal $\textrm{S}^7/(\mathbb{Z}_{rk}\times\mathbb{Z}_r)$ space. The second method is more systematic and leads to a full classification of 1/4 and 1/2 BPS WLs for general circular quivers with alternating Chen-Simons levels. It is based on directly imposing the invariance of the most general superconnection compatible with the symmetries of the theory under a given fraction of SUSY charges.

Following these strategies, we find 1/4 and 1/2 BPS WLs already constructed in the literature as the holonomy of double--node superconnections in \cite{Ouyang:2015qma,Cooke:2015ila} and \cite{Ouyang:2015bmy}, but also new ones that correspond to non--block--diagonal superconnections and are thus not ascribable to straightforward generalizations of the previous ones.
These new WLs are in general 1/4 BPS and they are enhanced to the usual 1/2 BPS for special choices of the parameters. As we will discuss in details, they are all cohomological equivalent to the bosonic 1/4 BPS operators whose expectation values can be in principle computed with localization techniques.

The paper is organized as follows. In section~\ref{ABJM} we review the classification of known BPS WLs in ABJ(M) theory by parametrizing the couplings to matter in the most general way.
In section~\ref{secorb} we determine the BPS WLs in the orbifold ABJM model that can be obtained by decomposition of BPS WLs in ABJM theory. In particular, we find new BPS WLs that were not present in the previous literature.
Section~\ref{secm2} is devoted to a discussion of the M2--/anti--M2--brane duals of BPS WLs in $\mN=4$ orbifold ABJM theory.
In section~\ref{secsusy}, by studying the invariance of the most general superconnection under a given fraction of SUSY transformations we give a complete classification of all possible 1/4 and 1/2 BPS WLs in $\mN=4$ circular quiver SCSM theories.
Finally, a summary of our main results is contained in section~\ref{seccon} where we also comment on the expected perturbative results for the newly found WLs and their matching with the localization prediction.
Technical details about the general derivation of section \ref{secsusy} are collected in appendix \ref{details}.
As a completion to the classification of BPS WLs in ABJ(M) and $\mathcal N=4$ SCSM theories, in appendix~\ref{secsup} we study the overlapping of preserved supercharges for general 1/2 BPS WLs in both theories.
We find no new pairs of different operators preserving the same set of supercharges beyond the ones already discovered in \cite{Cooke:2015ila} and \cite{Lietti:2017gtc}.
In the main body of the paper we focus on timelike linear WLs in Minkowski spacetime, whereas circular 1/4 BPS WLs in Euclidean space are discussed in appendix~\ref{appcir}.

\section{Review of BPS WLs in ABJ(M) theory}\label{ABJM}

 In this section we review the general classification of 1/6 and 1/2 BPS WLs in ABJ(M) theory given in \cite{Ouyang:2015iza,Ouyang:2015bmy}. We present it in a way that turns out to be preparatory for the study of BPS operators in $\mathcal N=4$ SCSM theories.

The ABJ(M) lagrangian in Minkowski spacetime can be written as the sum of four terms
\bea \label{lagrangian}
&& \hspace{-2mm}
   \mL_{CS} = \f{k}{4\pi}\ve^{\m\n\r}\Tr \Big( A_\m\p_\n A_\r +\f{2\ii}{3}A_\m A_\n A_\r
                                             -B_\m\p_\n B_\r -\f{2\ii}{3}B_\m B_\n B_\r  \Big) \nn\\
&& \hspace{-2mm}
   \mL_k = \Tr( -D_\m\bar\phi^I D^\m\phi_I + \ii\bar\psi_I\g^\m D_\m\psi^I ) \nn\\
&& \hspace{-2mm}
   \mL_p =\f{4\pi^2}{3k^2} \Tr(    \phi_I\bar\phi^I\phi_J\bar\phi^J\phi_K\bar\phi^K
                                +  \phi_I\bar\phi^J\phi_J\bar\phi^K\phi_K\bar\phi^I
                                +4 \phi_I\bar\phi^J\phi_K\bar\phi^I\phi_J\bar\phi^K
                                -6 \phi_I\bar\phi^J\phi_J\bar\phi^I\phi_K\bar\phi^K )  \nn\\
&& \hspace{-2mm}
   \mL_Y = \f{2\pi\ii}{k} \Tr(    \phi_I\bar\phi^I\psi^J\bar\psi_J
                               -2 \phi_I\bar\phi^J\psi^I\bar\psi_J
                               -  \bar\phi^I\phi_I\bar\psi_J\psi^J
                               +2 \bar\phi^I\phi_J\bar\psi_I\psi^J \nn\\
&& \hspace{-2mm}
   \phantom{\mL_Y = \f{2\pi\ii}{k} \Tr(}
                               +\ve^{IJKL}\phi_I\bar\psi_J\phi_K\bar\psi_L
                               -\ve_{IJKL}\bar\phi^I\psi^J\bar\phi^K\psi^L )
\eea
where the totally anti-symmetric Levi--Civita tensors $\ve^{IJKL}$, $\ve_{IJKL}$ are defined as $\ve^{1234}=\ve_{1234}=1$. Here $A_\mu$ and $B_\mu$ are the connections of $U(N)$ and $U(M)$ gauge groups respectively,  $\phi_I, \psi^I$ ($I=1,2,3,4$) are complex scalars and Dirac fermions in the bifundamental representation of the gauge group and in the fundamental representation of the $SU(4)$ R--symmetry group, and $\bar{\phi}^I, \bar{\psi}_I$ their complex conjugates. Covariant derivatives are defined as
\bea
&& D_\m\phi_I =\p_\m \phi_I + \ii A_\m \phi_I - \ii \phi_I B_\m \nn\\
&& D_\m\bar\phi^I =\p_\m \bar\phi^I - \ii \bar\phi^I A_\m + \ii B_\m \bar\phi^I  \nn\\
&& D_\m\psi^I =\p_\m \psi^I + \ii A_\m \psi^I - \ii \psi^I B_\m \nn \\
&& D_\m\bar\psi_I =\p_\m \bar\psi_I - \ii \bar\psi_I A_\m + \ii B_\m \bar\psi_I
\eea

The ABJ(M) action is invariant under the following Poincar\'e
SUSY transformations \cite{Gaiotto:2008cg,Hosomichi:2008jb,Terashima:2008sy,Bandres:2008ry}
\bea
&& \d A_\m=-\f{2\pi}{k} \lt( \phi_I\bar\psi_J\g_\m\th^{IJ} +\bar\th_{IJ}\g_\m\psi^J\bar\phi^I \rt) \nn\\
&& \d B_\m=-\f{2\pi}{k} \lt( \bar\psi_J\phi_I\g_\m\th^{IJ}+\bar\th_{IJ}\g_\m\bar\phi^I\psi^J \rt) \nn\\
&& \d\phi_I=\ii\bar\th_{IJ}\psi^J,   \qquad\d\bar\phi^I=\ii\bar\psi_J\th^{IJ} \nn\\
&& \d\psi^I=\g^\m\th^{IJ}D_\m\phi_J 
            +\f{2\pi}{k}\th^{IJ} \lt( \phi_J\bar\phi^K\phi_K-\phi_K\bar\phi^K\phi_J \rt)
            +\f{4\pi}{k}\th^{KL}\phi_K\bar\phi^I\phi_L \nn\\
&& \d\bar\psi_I=-\bar\th_{IJ}\g^\m D_\m\bar\phi^J 
                -\f{2\pi}{k}\bar\th_{IJ} \lt( \bar\phi^J\phi_K\bar\phi^K-\bar\phi^K\phi_K\bar\phi^J \rt)
                -\f{4\pi}{k}\bar\th_{KL}\bar\phi^K\phi_I\bar\phi^L
\eea
with the $\th^{IJ}$ parameters satisfying
\be
\th^{IJ}=-\th^{JI}, ~~ (\th^{IJ})^*=\bar \th_{IJ}, ~~ \bar\th_{IJ}=\f{1}{2}\ve_{IJKL}\th^{KL} 
\ee
${\cal N}=6$ SUSY is realized explicitly.

The whole spectrum of BPS WLs defined along the timelike infinite straight line $x^\m=(\t,0,0)$  can be described by a parametric family of operators \cite{Drukker:2009hy,Ouyang:2015iza,Ouyang:2015bmy}
\be \label{j69}
W = \mP \exp\Big( -\ii \int d\t L(\t) \Big)
\ee
corresponding to a generalized $U(N|M)$ superconnection that includes couplings to scalar and fermion matter fields\footnote{In three-dimensional Minkowski spacetime, we use the gamma matrix $\g^{\m~\b}_{~\a}=(\ii\s^2,\s^1,\s^3)$ with $\s^{1,2,3}$ being the Pauli matrices. For an arbitrary spinor $\th_\a$, we define $\th_\pm = \pm \ii u_\pm^\a \th_\a$ with the Grassmann even spinors $u_{\pm}^{\a} = \f{1}{\sqrt{2}}( \mp\ii, -1)$ \cite{Lee:2010hk,Lietti:2017gtc}.}
\be \label{l1}
L = \lt( \ba{cc} A_0 + \f{2\pi}{k} U^I_{~J} \phi_I\bar\phi^J &
         \sqrt{\f{4\pi}{k}}( \bar\a_I\psi^I_+ + \bar\g_I\psi^I_- ) \\
         \sqrt{\f{4\pi}{k}}( \bar\psi_{I-}\b^I - \bar\psi_{I+}\d^I ) &
         B_0 + \f{2\pi}{k} U^I_{~J} \bar\phi^J \phi_I \ea \rt)
\ee
Here $U^I_{~J}$ is a $4 \times 4$ matrix with constant complex entries, whereas the fermionic couplings are given by constant vectors $\bar\a_I$, $\bar{\gamma}_I$, $\beta^I$, $\delta^I$ in $\mathbb{C}^4$  ($\a^I=(\bar\a_I)^*$, $|\a|^2 = \bar\a_I\a^I$, etc.).

For special choices of the parameters, superconnection (\ref{l1}) gives rise to BPS WLs preserving a certain amount of SUSY, as we will review below for 1/6 and 1/2 BPS cases.

\subsection{1/6 BPS WLs} \label{ABJM1}

An exhaustive classification of 1/6 BPS WLs in ABJ(M) theory has been given in \cite{Ouyang:2015iza,Ouyang:2015bmy}. These are operators that preserve two real Poincar\'e supercharges plus two real superconformal charges out of the original $12 + 12$ real supercharges.

In order to make the classification clearer, it is convenient to define two projectors $P^I_{~J}$ and $Q^I_{~J}$ in the $SU(4)$ R--symmetry space that satisfy
\bea \label{identities}
&& P^I_{~J} + Q^I_{~J} = \d^I_J , ~~
   P^I_{~J}Q^J_{~K}=Q^I_{~J}P^J_{~K}=0 \nn\\
&& P^I_{~J}P^J_{~K}=P^I_{~K}, ~~
   Q^I_{~J}Q^J_{~K}=Q^I_{~K}, ~~
   P^I_{~I}=Q^I_{~I}=2
\eea
and break R--symmetry as $SU(4) \to SU(2)_L\times SU(2)_R$.
We can rewrite the two projectors as
\be \label{g4}
P^I_{~J} = \m^I \bar\m_J + \n^I \bar\n_J , ~~
Q^I_{~J} = \r^I \bar\r_J + \s^I \bar\s_J
\ee
in terms of four orthonormal vectors in $\mathbb{C}^4$ satisfying
\bea \label{constraints}
&& \bar\m_I\m^I = \bar\n_I\n^I = \bar\r_I\r^I = \bar\s_I\s^I = 1 \nn\\
&& \bar\m_I\n^I = \bar\m_I\r^I = \bar\m_I\s^I = \bar\n_I\r^I = \bar\n_I\s^I = \bar\r_I\s^I = 0
\eea
The two projectors are not independent, as $Q^I_{~J}$ can be expressed in terms of $P^I_{~J}$ using the first equation in (\ref{identities}). Moreover, for fixed $P^I_{~J}$ and $Q^I_{~J}$ there is some freedom in the choice of the $\bar\m_I$, $\bar\n_I$, $\bar\r_I$, $\bar\s_I$ vectors, which are always determined up to a $SU(2)_L\times SU(2)_R$ rotation.

\vskip 5pt
All 1/6 BPS WLs can be classified according to the choices of the parameters in table~\ref{tab1}. The $U^I_{~J}$ couplings are given in terms of the two projectors written as in (\ref{g4}) and we have expressed also the fermionic couplings as linear combination of the $\bar\m_I$, $\bar\n_I$, $\bar\r_I$, $\bar\s_I$ vectors  with four arbitrary constant complex parameters $p_I$. Their particular decompositions follow from the constraints $\bar\a_I P^I_{~J} = P^I_{~J} \b^J = \bar\g_I Q^I_{~J} = Q^I_{~J} \d^J = 0$ \cite{Ouyang:2015iza,Ouyang:2015bmy}.

In general, 1/6 BPS WLs in the first four classes include non--trivial couplings both to scalars and fermions.  For this reason they are called {\em fermionic} 1/6 BPS WLs, in contrast with $W^{\bos}_{1/6}$ that, including only couplings to scalars, is called {\em bosonic} 1/6 BPS WL. All the 1/6 BPS WLs preserve the same set of Poincar\'e supercharges\footnote{For WLs on an infinite straight line, Poincar\'e and conformal supercharges are separately and similarly preserved. Therefore, it is sufficient to discuss WLs invariance under Poincar\'e supercharges.}, which correspond to
\be \label{g1}
P^I_{~K}P^J_{~L}\th^{KL}_-, ~~ Q^I_{~K}Q^J_{~L}\th^{KL}_+
\ee

\begin{table}[htbp]
\centering
\begin{tabular}{|c|c|}\hline
  WL type & Choice of the parameters \\ \hline\hline

  \multirow{2}*{$W_{1/6}^\I[\bar\m_I,\bar\n_I,p_I]$}
  & $U^I_{~J} = \m^I \bar\m_J + \n^I \bar\n_J - (1-2\bar\a_K\b^K) (\d^I_J - \m^I \bar\m_J - \n^I \bar\n_J) - 2\b^I\bar\a_J $ \\
  & $\bar\a_I = p_1 \bar\r_I + p_2 \bar\s_I, ~~
     \b^I = p_3 \r^I + p_4 \s^I, ~~
     \bar\g_I = \d^I = 0$ \\ \hline

  \multirow{2}*{$W_{1/6}^\II[\bar\m_I,\bar\n_I,p_I]$}
  & $U^I_{~J} = (1-2\bar\g_K\d^K) (\m^I \bar\m_J + \n^I \bar\n_J) + 2\d^I\bar\g_J - (\d^I_J - \m^I \bar\m_J - \n^I \bar\n_J)$ \\
  & $\bar\g_I = p_1 \bar\m_I + p_2 \bar\n_I, ~~
     \d^I = p_3 \m^I + p_4 \n^I, ~~
     \bar\a_I = \b^I = 0$ \\ \hline

  \multirow{2}*{$W_{1/6}^\III[\bar\m_I,\bar\n_I,p_I]$}
  & $U^I_{~J} = - \d^I_J + 2\m^I \bar\m_J + 2\n^I \bar\n_J $ \\
  & $\bar\a_I = p_1 \bar\r_I + p_2 \bar\s_I, ~~
     \bar\g_I = p_3 \bar\m_I + p_4 \bar\n_I, ~~
     \b^I = \d^I = 0$ \\ \hline

  \multirow{2}*{$W_{1/6}^\IV[\bar\m_I,\bar\n_I,p_I]$}
  & $U^I_{~J} =  - \d^I_J + 2\m^I \bar\m_J + 2\n^I \bar\n_J $ \\
  & $\b^I = p_1 \r^I + p_2 \s^I, ~~
     \d^I = p_3 \m^I + p_4 \n^I, ~~
     \bar\a_I = \bar\g_I = 0$ \\ \hline \hline

  $W^{\bos}_{1/6}[\bar\m_I,\bar\n_I]$
  & $U^I_{~J} = - \d^I_J + 2\m^I \bar\m_J + 2\n^I \bar\n_J , ~~
     \bar\a_I = \bar\g_I = \b^I = \d^I =0$ \\ \hline

\end{tabular}
\caption{The four types of {\em fermionic} WLs and the {\em bosonic} 1/6 BPS WL in ABJ(M) theory. In writing the $U^I_{~J}$ matrices we have used $\r^I \bar\r_J + \s^I \bar\s_J=  \d^I_J - \m^I \bar\m_J - \n^I \bar\n_J$, which follows from the first identity in (\ref{identities}).}\label{tab1}
\end{table}

Different classes of WLs give rise to independent operators, as they cannot be mapped one into the other by R--symmetry rotations. In fact, $W_{1/6}^\I$ and $W_{1/6}^\II$ operators break the original $SU(4)$ R--symmetry down to $SU(2)_L$ and $SU(2)_R$, respectively.
For $W_{1/6}^\III$ and $W_{1/6}^\IV$, the R--symmetry is broken completely.
Finally, $W^{\bos}_{1/6}$ is invariant under the $SU(2)_L \times SU(2)_R$ subgroup.

After fixing $P^I_{~J}$, $Q^I_{~J}$ in the R--symmetry space, i.e., fixing the preserved supercharges (\ref{g1}), each of the four types of fermionic WLs depends on four independent complex parameters $p_I$, while the bosonic WL is totally fixed. The bosonic WL can be obtained from any of the four fermionic WLs by setting $p_I=0$, $I=1,2,3,4$, and so it is just one particular representative of the four families of 1/6 BPS WLs.

It is important to recall that all the fermionic 1/6 BPS WLs are cohomologically equivalent to the bosonic 1/6 BPS WL, being expressible as \cite{Drukker:2009hy,Ouyang:2015bmy}
\be \label{cohom}
W_{1/6}^{\I,\II,\III,\IV} [\bar\m_I,\bar\n_I,p_I] = W_{1/6}^{\bos}[\bar\m_I,\bar\n_I] + {\mQ} V^{\I,\II,\III,\IV}[\bar\m_I,\bar\n_I,p_I]
\ee
where $\mQ$ is a linear combination of preserved supercharges (\ref{g1}). This property turns out to be very important when computing vacuum expectation values (vev) of Euclidean circular BPS WLs in the ABJ(M) theory compactified on S$^3$. In fact, using the ${\cal Q}$ supercharge in (\ref{cohom}) to localize the path integral, the cohomological relation   implies that at quantum level all  the vev are identical and equal to $\langle W_{1/6}^{\bos} \rangle$ computed by a matrix model \cite{Kapustin:2009kz,Marino:2009jd,Drukker:2010nc}.

\subsection{1/2 BPS WLs} \label{ABJM2}

For special values of the parameters in table \ref{tab1}, SUSY gets enhanced and we obtain operators that preserve half of the supersymmetries \cite{Drukker:2009hy}. Precisely, this happens in Class I by setting $\b^I = \f{\a^I}{|\a|^2}$, $|\a|^2\neq0$ and in Class II for $\d^I = \f{\g^I}{|\g|^2}$, $|\g|^2\neq0$. In our notations of table \ref{tab1} this corresponds to choosing
\be \label{choice}
p_3=\f{\bar p_1}{p_1\bar p_1 + p_2\bar p_2} \qquad p_4=\f{\bar p_2}{p_1\bar p_1 + p_2\bar p_2} \qquad p_1\bar p_1 + p_2\bar p_2 \neq 0
\ee
We denote the corresponding 1/2 BPS WLs as $W^\I_{1/2}[\bar\a_I]$ and  $W^\II_{1/2}[\bar\g_I]$, respectively. They are manifestly invariant under a residual $SU(3)$ R--symmetry subgroup.

Being particular cases of $W_{1/6}^{\I,\II}[\bar\m_I,\bar\n_I,p_I]$ families, these operators have to preserve supercharges (\ref{g1}). However, due to their $SU(3)$ invariance, the corresponding sets of preserved supercharges contain also supercharges that are $SU(3)$ rotations of (\ref{g1}). It is easy to prove that this enlarges the set of preserved SUSY's to
\bea \label{preserved}
&& W^\I_{1/2}[\bar\a_I]: \quad \bar\a_J\th^{IJ}_+, ~~ \ve_{IJKL}\a^J \th^{KL}_- \nn \\
&& W^\II_{1/2}[\bar\g_I]: \quad  \bar\g_J\th^{IJ}_-, ~~
  \ve_{IJKL}\g^J \th^{KL}_+
\eea

As long as we consider the two 1/2 BPS WLs as particular representatives of Class I and Class II fermionic WLs, the corresponding parameters are necessarily orthogonal, i.e. $\bar{\a}_I \g^I= 0$. However, at this point nothing prevents from relaxing these conditions and freely rotating the two vectors in the $SU(4)$ R--symmetry space. From (\ref{preserved}) it then follows that for a given representative of Class $\I$ selected by choosing a specific vector $\bar\a_I$ in $\mathbb{C}^4$, it is possible to select a representative in Class $\II$ corresponding to $\bar\g_I = \bar\a_I$ that preserves the complementary set of supercharges. For ABJM theory, they are in fact dual to a pair of M2--/anti--M2--branes placed at the same position in $\gabjm$ spacetime \cite{Lietti:2017gtc}. For a more general discussion on the overlapping of preserved supercharges for different values of the parameters we refer to appendix \ref{secsup}.

From identity (\ref{cohom}) and R--symmetry, it follows that $W^{\I}_{1/2}[\bar\a_I]$ is cohomologically equivalent to the bosonic 1/6 BPS WL $W_{1/6}^\bos[\bar\m_I,\bar\n_I]$ for arbitrary $\bar\m_I$, $\bar\n_I$ satisfying $\bar\a_I\m^I=\bar\a_I\n^I=\bar\m_I\n^I=0$, $\bar\m_I\m^I=\bar\n_I\n^I=1$. Analogously, $W^\II_{1/2}[\bar\g_I]$ is cohomologically equivalent to the bosonic 1/6 BPS WL $W_{1/6}^\bos[\bar\g_I/|\g|^2,\bar\n_I]$ for arbitrary $\bar\n_I$ that satisfies $\bar\g_I\n^I=0$, $\bar\n_I\n^I=1$.

\section{BPS WLs in $\mN=4$ orbifold ABJM theory}\label{secorb}

 As is well--known, a particular realization of Chern--Simons--matter theory with $\mN=4$ supersymmetry can be obtained by orbifold projection of the ABJM theory \cite{Benna:2008zy}. Precisely, starting from the ABJM theory with gauge group and levels $U(rN)_k \times U(rN)_{-k}$ and taking a $\mathbb{Z}_r$ quotient one obtains the $\mN=4$ orbifold ABJM theory corresponding to a circular quiver $[U(N)_k \times U(N)_{-k}]^r$ with alternating levels.
 
The study of BPS WLs in orbifold ABJM theory was initiated in \cite{Ouyang:2015qma,Cooke:2015ila} and refined in \cite{Ouyang:2015bmy}, both for straight line contours in Minkowski and circular contours in Euclidean space.  The presently known classification includes one bosonic and four classes of fermionic WLs obtained by assuming that the corresponding superconnection can be always written as a $2 \times 2$ block diagonal matrix. As a consequence, when traced, the WL can be written as a linear combination of WLs connecting two adjacent quiver nodes.

Aimed at extending this classification, here we use an alternative approach to construct WLs in orbifold ABJM that does not require making any particular ansatz on the structure of the superconnection. Precisely, we construct a class of Wilson operators by performing the orbifold decomposition of WLs of the ABJM theory.

The orbifold decomposition breaks the original $SU(4)$ R--symmetry to $SU(2)_L\times SU(2)_R$.  Correspondingly, we choose the following decomposition of the R--symmetry indices
\be \label{indices}
I = 1,2,4,3 ~\to~ i=1,2, ~ \hi=\hat 1, \hat 2
\ee
A generic vector $\bar\a_I$ in $\mathbb{C}^4$ is then decomposed as
\be \label{vector}
\bar\a_I ~\to~ \bar\a_i, \bar\a_\hi
\ee
with complex conjugates $\a^i=(\bar\a_i)^*, \a^\hi=(\bar\a_\hi)^*$. We also define $|\a|^2 = \bar\a_i\a^i + \bar\a_\hi\a^\hi$.

 In the $U(rN)_k \times U(rN)_{-k}$ ABJM model we consider the most general superconnection of the form (\ref{l1}). Applying the orbifold decomposition the supermatrix gets decomposed
for $r \geq 3$ as
\be \label{l2}
L = \lt(\ba{cccccccc}
\mA^{(1)}    & f_1^{(1)}  & h_1^{(1)} &   0        &     0         &    \cdots          & h_2^{(2r-1)} & f_2^{(2r)} \\
f_2^{(1)}    & \mB^{(2)}  & f_1^{(2)} & h_1^{(2)} &     0         &    \cdots          &       0       & h_2^{(2r)} \\
h_2^{(1)}    & f_2^{(2)}  & \mA^{(3)} & f_1^{(3)} & \ddots       &           &         0     &  0\\
      0       & h_2^{(2)}  & f_2^{(3)} & \mB^{(4)} & \ddots       & \ddots       &              & \vdots\\
    0         &            & \ddots    & \ddots    & \ddots       & \ddots       & h_1^{(2r-3)} & 0 \\
    \vdots         &            &           & \ddots    & \ddots       & \ddots       & f_1^{(2r-2)} & h_1^{(2r-2)} \\
h_1^{(2r-1)} &    0        &   \cdots        &       0    & h_2^{(2r-3)} & f_2^{(2r-2)} & \mA^{(2r-1)} & f_1^{(2r-1)} \\
f_1^{(2r)}   & h_1^{(2r)} &     0      &    \cdots       &      0        & h_2^{(2r-2)} & f_2^{(2r-1)} & \mB^{(2r)}
\ea\rt)
\ee
and for $r=2$ as
\be \label{l3}
L = \lt(\ba{cccc}
\mA^{(1)}           & f_1^{(1)}           & h_1^{(1)}+h_2^{(3)} & f_2^{(4)}           \\
f_2^{(1)}           & \mB^{(2)}           & f_1^{(2)}           & h_1^{(2)}+h_2^{(4)} \\
h_1^{(3)}+h_2^{(1)} & f_2^{(2)}           & \mA^{(3)}           & f_1^{(3)} \\
f_1^{(4)}           & h_1^{(4)}+h_2^{(2)} & f_2^{(3)}           & \mB^{(4)} \\
\ea\rt)
\ee
where we have defined
\bea
 \label{entries}
&& \mA^{(2\ell-1)} =  A_0^{(2\ell-1)}
   + \f{2\pi}{k} \Big( U^i_{~j} \phi_i^{(2\ell-2)}\bar\phi^j_{(2\ell-2)}
                      +U^\hi_{~\hj} \phi_\hi^{(2\ell-1)}\bar\phi^\hj_{(2\ell-1)}
                 \Big) \nn\\
&& \mB^{(2\ell)} =  B_0^{(2\ell)}
   + \f{2\pi}{k} \Big( U^i_{~j} \bar\phi^j_{(2\ell)}\phi_i^{(2\ell)}
                      +U^\hi_{~\hj}\bar\phi^\hj_{(2\ell-1)}\phi_\hi^{(2\ell-1)}
                 \Big) \nn\\ \label{entries}
&& f_1^{(2\ell-1)} = \sqrt{\f{4\pi}{k}} \Big( \bar\a_i\psi^i_{(2\ell-1)+} + \bar\g_i\psi^i_{(2\ell-1)-} \Big), \qquad
   f_1^{(2\ell)} = \sqrt{\f{4\pi}{k}} \Big( \bar \psi_{\hi-}^{(2\ell)} \b^\hi - \bar \psi_{\hi+}^{(2\ell)} \d^\hi \Big) \nn\\
&& f_2^{(2\ell-1)} = \sqrt{\f{4\pi}{k}} \Big( \bar \psi_{i-}^{(2\ell-1)} \b^i - \bar \psi_{i+}^{(2\ell-1)} \d^i \Big) , \qquad \quad
   f_2^{(2\ell)} = \sqrt{\f{4\pi}{k}} \Big( \bar\a_\hi\psi^\hi_{(2\ell)+} + \bar\g_\hi\psi^\hi_{(2\ell)-} \Big) \nn\\
&& h_1^{(2\ell-1)} = \f{2\pi}{k} U^\hi_{~j} \phi_\hi^{(2\ell-1)}\bar\phi^j_{(2\ell)}, \qquad \qquad \qquad
   h_1^{(2\ell)} = \f{2\pi}{k} U^\hi_{~j} \bar\phi^j_{(2\ell)}\phi_\hi^{(2\ell+1)}\nn\\
&& h_2^{(2\ell-1)} = \f{2\pi}{k} U^i_{~\hj} \phi_i^{(2\ell)}\bar\phi^\hj_{(2\ell-1)}, \qquad \qquad \qquad
   h_2^{(2\ell)} = \f{2\pi}{k} U^i_{~\hj} \bar\phi^\hj_{(2\ell+1)}\phi_i^{(2\ell)}
\eea
In superconnections (\ref{l2}) and (\ref{l3}) we have the usual  gauge and scalar field couplings in the diagonal blocks, fermion fields in the next-to-diagonal blocks, and also new scalar field couplings in the next-to-next-to-diagonal blocks.  The novelty here is the presence of these next-to-next-to-diagonal blocks.\footnote{Non--vanishing next-to-next-to-diagonal blocks have been considered in \cite{Cooke:2015ila} only for 1/2 BPS WLs in $\mN=4$ SCSM theories with some quiver nodes corresponding to vanishing CS levels. }. These new blocks, present both in the bosonic and fermionic WLs, together with generically non--vanishing $f_{1,2}^{(2\ell-1)}$ and $f_{1,2}^{(2\ell)}$ blocks define superconnections that are outside the general class of $2 \times 2$ block diagonal superconnections considered so far in the literature  \cite{Ouyang:2015qma,Cooke:2015ila,Ouyang:2015hta,Ouyang:2015bmy}. They reduce to those ones for the particular choice of the couplings  $U^\hi_{~j} = U^i_{~\hj}  = 0$ and either $f_{1,2}^{(2\ell-1)}=0$ or $f_{1,2}^{(2\ell)}=0, \forall \, \ell$.

In the next two subsections we give explicit examples of the construction of BPS WLs in orbifold ABJM theory from orbifold reduction of 1/6 and 1/2 BPS operators in ABJM theory.

\subsection{From 1/6 BPS WLs in ABJM theory}

We begin by considering the decomposition of the BPS WLs reviewed in subsection \ref{ABJM1}.
Decomposing the constant vector parameters of the general superconnection (\ref{l1}) according to (\ref{vector}) and imposing constraints (\ref{constraints})
we obtain generic WLs $W[\bar\m_i, \bar\m_\hi,\bar\n_i,\bar\n_\hi, p_I] $ with
\be
 \bar\m_i\m^i + \bar\m_\hi\m^\hi = \bar\n_i\n^i + \bar\n_\hi\n^\hi  = 1, \qquad
\bar\m_i\n^i + \bar\m_\hi\n^\hi =0
\ee
and $p_I$ still labeling four constant complex numbers.

Imposing that the new operators in the $\mN=4$ theory preserve some amount of SUSY provides further constraints on the complex parameters. We find that the solution $\bar\m_\hi = 0$, $\bar\n_i =0$ leads to fermionic BPS WLs preserving two Poincar\'e and two conformal supercharges.\footnote{An alternative solution would be $\bar\m_i = 0$, $\bar\n_\hi =0$, which leads to equivalent WLs. Without loss of generality we discuss only one case.}
The resulting operators, obtained from the four classes of fermionic WLs in table \ref{tab1}, and the corresponding parameters are classified in table \ref{tab2}.

It can be shown that all the four types of fermionic BPS WLs preserve the same set of supercharges corresponding to
\be \label{e15}
  \bar\m_i\bar\n_\hj\th^{i\hj}_-, ~~ \ve_{ij}\ve_{\hk\hl}\m^i\n^\hk\th^{j\hl}_+
  \ee
The important observation is that the classes of fermionic 1/4 BPS WLs found here through the orbifolding projection are more general than the ones constructed in \cite{Ouyang:2015bmy}. In fact, for generic parameters superconnections (\ref{l2}), (\ref{l3}) are not block--diagonal and cannot be mapped simply by R--symmetry rotations to the block--diagonal matrices previously considered in the literature \cite{Ouyang:2015qma,Cooke:2015ila,Ouyang:2015hta,Ouyang:2015bmy,Lietti:2017gtc}.

\begin{table}[htbp]
  \centering
  \begin{tabular}{|c|c|}\hline

  WL type & Choice of the parameters \\ \hline\hline

  \multirow{4}*{$W_{1/4}^\I[\bar\m_i,0,0,\bar\n_\hi,p_I]$}
  & $U^i_{~j} = \m^i\bar\m_j - ( 1-2\bar\a_k\b^k -2 \bar\a_\hk\b^\hk ) ( \d^i_j - \m^i\bar\m_j ) - 2\b^i\bar\a_j$  \\
  & $U^\hi_{~\hj} = \n^\hi\bar\n_\hj - ( 1-2\bar\a_k\b^k -2 \bar\a_\hk\b^\hk ) ( \d^\hi_\hj - \n^\hi\bar\n_\hj ) - 2\b^\hi\bar\a_\hj$ \\
  & $U^i_{~\hj} = - 2\b^i\bar\a_\hj, ~~
     U^\hi_{~j} = - 2\b^\hi\bar\a_j, ~~
     \bar\g_i=\bar\g_\hi=\d^i=\d^\hi=0$ \\
  & $\bar\a_i=p_1\ve_{ij}\m^j, ~~
     \bar\a_\hi=p_2\ve_{\hi\hj}\n^\hj, ~~
     \b^i=p_3\ve^{ij}\bar\m_j, ~~
     \b^\hi=p_4\ve^{\hi\hj}\bar\n_\hj$ \\ \hline

  \multirow{4}*{$W_{1/4}^\II[\bar\m_i,0,0,\bar\n_\hi,p_I]$}
  & $U^i_{~j} = ( 1-2\bar\g_k\d^k -2 \bar\g_\hk\d^\hk )\m^i\bar\m_j + 2\d^i\bar\g_j - ( \d^i_j - \m^i\bar\m_j )$   \\
  & $U^\hi_{~\hj} = ( 1-2\bar\g_k\d^k -2 \bar\g_\hk\d^\hk )\n^\hi\bar\n_\hj + 2\d^\hi\bar\g_\hj - ( \d^\hi_\hj - \n^\hi\bar\n_\hj )$ \\
  & $U^i_{~\hj} = 2\d^i\bar\g_\hj, ~~
     U^\hi_{~j} = 2\d^\hi\bar\g_j, ~~
     \bar\a_i=\bar\a_\hi=\b^i=\b^\hi=0$ \\
  & $\bar\g_i=p_1\bar\m_i, ~~
     \bar\g_\hi=p_2\bar\n_\hi, ~~
     \d^i=p_3 \m^i$, ~~ $\d^\hi=p_4\n^\hi$ \\ \hline

  \multirow{3}*{$W_{1/4}^\III[\bar\m_i,0,0,\bar\n_\hi,p_I]$}
  & $U^i_{~j} = - \d^i_j  + 2\m^i\bar\m_j , ~~
     U^\hi_{~\hj} = - \d^\hi_\hj + 2 \n^\hi\bar\n_\hj$  \\
  & $U^i_{~\hj} = U^\hi_{~j} = \b^i=\b^\hi=\d^i=\d^\hi=0$ \\
  & $\bar\a_i=p_1\ve_{ij}\m^j, ~~
     \bar\a_\hi=p_2\ve_{\hi\hj}\n^\hj, ~~
     \bar\g_i=p_3\bar\m_i, ~~
     \bar\g_\hi=p_4\bar\n_\hi$ \\ \hline

  \multirow{3}*{$W_{1/4}^\IV[\bar\m_i,0,0,\bar\n_\hi,p_I]$}
  & $U^i_{~j} = - \d^i_j  + 2\m^i\bar\m_j , ~~
     U^\hi_{~\hj} = - \d^\hi_\hj + 2 \n^\hi\bar\n_\hj$   \\
  & $U^i_{~\hj} = U^\hi_{~j} = \bar\a_i = \bar\a_\hi = \bar\g_i = \bar\g_\hi = 0$ \\
  & $\b^i=p_1\ve^{ij}\bar\m_j, ~~
     \b^\hi=p_2\ve^{\hi\hj}\bar\n_\hj, ~~
     \d^i=p_3 \m^i, ~~
     \d^\hi=p_4\n^\hi$\\ \hline \hline

  \multirow{2}*{$W^{\bos}_{1/4}[\bar\m_i,0,0,\bar\n_\hi]$}
  & $U^i_{~j} = -\d^i_j + 2 \m^i\bar\m_j, ~~ U^\hi_{~\hj} = -\d^\hi_\hj + 2 \n^\hi\bar\n_\hj$ \\
  & $U^i_{~\hj} = U^\hi_{~j}=\bar\a_i = \bar\a_\hi = \bar\g_i = \bar\g_\hi =
    \b^i=\b^\hi=\d^i=\d^\hi=0$ \\ \hline
  \end{tabular}
\caption{The four types of {\em fermionic} WLs and the {\em bosonic} 1/4 BPS WL in circular quiver $\mN=4$ SCSM theories with alternating levels. We have set $\bar\m_\hi=\bar\n_i=0$, and thus $\bar\m_i\m^i=\bar\n_\hi\n^\hi=1$.} \label{tab2}
\end{table}

In order to better clarify this point, we focus on particular WL representatives in each class selected by  choosing for instance  $\bar\m_i = (0,1)$ and $\bar\n_\hi = (0,1)$. As long as we keep the four $p_I$ parameters generically different from zero, from (\ref{entries}) we read that superconnections in Class I have non--trivial entries
\bea
& f_1^{(2\ell-1)} = p_1 \, \sqrt{\df{4\pi}{k}} \psi^1_{(2\ell-1)+} \qquad\qquad
& f_1^{(2\ell)} = p_4 \, \sqrt{\df{4\pi}{k}} \bar \psi_{\ho-}^{(2\ell)} \nn\\
& f_2^{(2\ell-1)} = p_3 \, \sqrt{\df{4\pi}{k}} \bar \psi_{1-}^{(2\ell-1)} \qquad\qquad
& f_2^{(2\ell)} = p_2 \, \sqrt{\df{4\pi}{k}}\psi^\ho_{(2\ell)+} \nn\\
& h_{1}^{(2\ell-1)} = - \dfrac{4\pi}{k} \, p_1p_4 \, \phi_{\hat{1}}^{(2\ell - 1)} \bar{\phi}^1_{(2\ell)} \qquad \qquad
& h_{1}^{(2\ell)} = - \dfrac{4\pi}{k} \, p_1p_4 \, \bar{\phi}^1_{(2\ell)} \phi_{\hat{1}}^{(2\ell + 1)}\nn \\
& h_{2}^{(2\ell-1)} = - \dfrac{4\pi}{k} \, p_2p_3 \, \phi_{1}^{(2\ell)} \bar{\phi}^{\hat{1}}_{(2\ell - 1)} \qquad \qquad
& h_{2}^{(2\ell)} = - \dfrac{4\pi}{k} \, p_2p_3 \, \bar{\phi}^{\hat{1}}_{(2\ell + 1)} \phi_{1}^{(2\ell)}
\eea
that prevent the superconnections to be written as block diagonal matrices. For Class II we find a similar pattern.
Although operators in Class III and IV always have $h_{1,2}^{(2\ell-1)} = h_{1,2}^{(2\ell)} =0$, still they have non--vanishing $f_1^{(2\ell-1)}$, $f_{2}^{(2\ell)}$ and $f_{2}^{(2\ell-1)}$, $f_1^{(2\ell)}$ respectively,
\bea
&& \hspace{-3mm} {\rm Class ~III:} ~
   f_1^{(2\ell-1)} = \sqrt{\f{4\pi}{k}} \Big( p_1 \, \psi^1_{(2\ell-1)+} + p_3 \, \psi^2_{(2\ell-1)-} \Big), ~~
   f_2^{(2\ell)} = \sqrt{\f{4\pi}{k}} \Big( p_2 \, \psi^\ho_{(2\ell)+} + p_4 \, \psi^\hw_{(2\ell)-} \Big) \nn\\
&& \hspace{-3mm} {\rm Class ~IV:} ~
   f_2^{(2\ell-1)} = \sqrt{\f{4\pi}{k}} \Big( p_1 \, \bar \psi_{1-}^{(2\ell-1)} - p_3 \, \bar \psi_{2+}^{(2\ell-1)} \Big) , ~~ \quad
   f_1^{(2\ell)} = \sqrt{\f{4\pi}{k}} \Big( p_2 \, \bar \psi_{\ho-}^{(2\ell)} - p_4 \, \bar \psi_{\hw+}^{(2\ell)} \Big) \nonumber \\
\eea
which do not allow to make them block--diagonal.

However, it can be easily realized that by choosing $p_2 = p_4 = 0$ we have $h_{1,2}^{(2\ell-1)} = h_{1,2}^{(2\ell)} = f_{1,2}^{(2\ell)}=0$, and in all the cases the superconnections become block--diagonal
\be
L = {\rm diag}(L_1, L_2, \dots, L_r)~, \qquad
L_\ell =  \lt(\ba{cc}
\mA^{(2\ell-1)} & f_1^{(2\ell-1)}                      \\
f_2^{(2\ell-1)} & \mB^{(2\ell)}            \\
\ea\rt)
\ee
The corresponding traced WL
\be
{\rm Tr} \, \mP \exp\Big( -\ii \int d\t L(\t) \Big)
\ee
can then be written as a linear combination of double--node operators $W^{(\ell)}$, which are the holonomy of superconnections $L_\ell$ with $\ell=1,2,\cdots,r$.

Similarly, by choosing $p_1 = p_3 = 0$, in all the cases we have $h_{1,2}^{(2\ell-1)} = h_{1,2}^{(2\ell)} = f_{1,2}^{(2\ell-1)}=0$ and the superconnections become block--diagonal
\be
\td{L} = {\rm diag}(\td{L}_1, \td{L}_2, \dots, \td{L}_r), ~~
\td{L}_\ell =  \lt(\ba{cc}
\mB^{(2\ell)} & f_1^{(2\ell)}                      \\
f_2^{(2\ell)} & \mA^{(2\ell+1)}\\
\ea\rt)
\ee
Fermionic 1/4 BPS WLs with block--diagonal superconnections have been considered in \cite{Ouyang:2015bmy}.
Our classification generalizes the previous one, providing the previously found 1/4 BPS WLs but also an infinite set of new 1/4 BPS operators.

To complete this picture we still need to consider the decomposition of the bosonic 1/6 BPS WL $W_{1/6}^{\bos}[\bar\m_I,\bar\n_I]$ in table \ref{tab1}. We obtain an operator $W^{\bos}_{1/4}[\bar\m_i,0,0,\bar\n_\hi]$ with superconnections (\ref{l2}) or (\ref{l3}) and parameters given in table \ref{tab2}. It is a bosonic 1/4 BPS WL that preserves supercharges (\ref{e15}). It can also be got from $W_{1/4}^{\I,\II,\III,\IV}[\bar\m_i,0,0,\bar\n_\hi,p_I]$ by setting $p_I=0$. The bosonic 1/4 BPS WL has been constructed in \cite{Ouyang:2015qma,Cooke:2015ila} and coincides with the present ones up to a R--symmetry rotation.

Finally, it is easy to see that cohomological equivalence (\ref{cohom}) survives the decomposition. Therefore, fermionic 1/4 BPS WLs in table \ref{tab2} are equivalent to the bosonic 1/4 BPS WL up to a ${\cal Q}$--exact term
\be
W_{1/4}^{\I,\II,\III,\IV}[\bar\m_i,0,0,\bar\n_\hi,p_I] = W_{1/4}^\bos[\bar\m_i,0,0,\bar\n_\hi] + \mQ V^{\I,\II,\III,\IV}[\bar\m_i,0,0,\bar\n_\hi,p_I]
\ee
Therefore, their vev computed on the three sphere are expected to coincide.

\vskip 10pt
\subsection{From 1/2 BPS WLs in ABJM theory}\label{sec3.2}

As a particular case of the previous analysis we now consider the decomposition of the 1/2 BPS WLs $W^\I_{1/2}[\bar\a_I]$, $W^\II_{1/2}[\bar\g_I]$ in ABJM theory. This turns out to be particularly helpful for studying possible enhancement of SUSY for operators in table \ref{tab2}, and for finding out the cases for which gravity duals are known, as we discuss in the next section.

Decomposing $W^\I_{1/2}[\bar\a_I]$ in table \ref{tab1}, we obtain an operator $W^\I[\bar\a_i,\bar\a_\hi]$ with superconnection (\ref{l2}) or (\ref{l3}) and parameters
\bea \label{1/2WL}
&& U^i_{~j} = \d^i_j - \f{2\a^i\bar\a_j}{|\a|^2}, ~~
U^\hi_{~\hj} = \d^\hi_\hj - \f{2\a^\hi\bar\a_\hj}{|\a|^2}, ~~
U^i_{~\hj} = - \f{2\a^i\bar\a_\hj}{|\a|^2} , ~~U^\hi_{~j} = - \f{2\a^\hi\bar\a_j}{|\a|^2} ~~\nn\\ 
&&
\b^i = \f{\a^i}{|\a|^2}, ~~
\b^\hi = \f{\a^\hi}{|\a|^2}, ~~
|\a|^2=\bar\a_i\a^i+\bar\a_\hi\a^\hi \neq 0, ~~
\bar\g_i=\bar\g_\hi=\d^i=\d^\hi=0
\eea
This operator can be seen as a special case of the fermionic 1/4 BPS WLs $W_{1/4}^\I[\bar\m_i,0,0,\bar\n_\hi,p_I]$ in table \ref{tab2}, obtained by choosing the parameters
as in (\ref{choice}). It can be 1/2 or 1/4 BPS depending on the parameters. In fact, we can distinguish three cases.
\begin{enumerate}

  \item[1)] When $\bar\a_i\a^i\neq0$, $\bar\a_\hi = 0$, we obtain a 1/2 BPS WL $W^\I_{1/2}[\bar\a_i,0]$, with preserved supercharges
  \be \label{e31}
  \bar\a_i \th^{i\hk}_+, ~~ \ve_{ij}\a^i\th^{j\hk}_- \qquad\hk = \ho,\hw
  \ee
as follows from (\ref{preserved}). For $\bar\a_i=\d_i^1$ it coincides with the $\psi_1$-loop in \cite{Ouyang:2015qma,Cooke:2015ila}, alternatively called $W_1$ in \cite{Lietti:2017gtc}, whereas for $\bar\a_i=\d_i^2$ it is the $W_2$ operator of \cite{Lietti:2017gtc}. It is cohomologically equivalent to the bosonic 1/4 BPS WL
\be
W_{1/4}^\bos[\ve_{ij}\a^j/\sqrt{\bar\a_k\a^k},0,0,\bar\n_\hi]
\ee
for arbitrary $\bar\n_\hi$, with $\bar\n_\hi\n^\hi=1$.

  \item[2)] When $\bar\a_i = 0$, $\bar\a_\hi\a^\hi\neq0$, we have the 1/2 BPS WL $W^\I_{1/2}[0,\bar\a_\hi]$ with preserved supercharges
  \be \label{e32} \bar\a_\hj \th^{i\hj}_+, ~~ \ve_{\hj\hk}\a^\hj\th^{i\hk}_- \qquad i=1,2 \ee
In \cite{Lietti:2017gtc} this operator was called $W_\ho$ for $\bar\a_\hi=\d_\hi^\ho$ and $W_\hw$ for $\bar\a_\hi=\d_\hi^\hw$. It is cohomologically equivalent to the bosonic 1/4 BPS WL \be W_{1/4}^\bos[\bar\m_i,0,0,\ve_{\hi\hj}\a^\hj/\sqrt{\bar\a_\hk\a^\hk}] \ee for arbitrary $\bar\m_i$ with $\bar\m_i\m^i=1$.

  \item[3)] More interestingly, when $\bar\a_i\a^i\neq0$ and $\bar\a_\hi\a^\hi\neq0$, we have the 1/4 BPS WL $W_{1/4}^\I[\bar\a_i,\bar\a_\hi]$ with preserved supercharges
  \be \label{e33}
  \bar\a_i\bar\a_\hj\th^{i\hj}_+, ~~ \ve_{ij}\ve_{\hk\hl}\a^i\a^\hk\th^{j\hl}_-
  \ee
  It is cohomologically equivalent to the bosonic 1/4 BPS WL
  \be
  W_{1/4}^\bos[\ve_{ij}\a^j/\sqrt{\bar\a_k\a^k},0,0, \ve_{\hi\hj}\a^\hj/\sqrt{\bar\a_\hk\a^\hk}]
  \ee

\end{enumerate}

Repeating a similar analysis for the 1/2 BPS WL $W^\II_{1/2}[\bar\g_I]$ in ABJM theory we obtain an operator $W^\II[\bar\g_i,\bar\g_\hi]$ in $\mN=4$ SCSM theories.  This is just a particular case of the fermionic 1/4 BPS WL $W_{1/4}^\II[\bar\m_i,0,0,\bar\n_\hi,p_I]$ in table~\ref{tab2} corresponding to parameters (\ref{choice}).
When $\bar\g_\hi=0$, $\bar\g_i\g^i\neq0$ we have a 1/2 BPS WL, $W^\II_{1/2}[\bar\g_i,0]$, which for $\bar\g_i = \d_i^2$ coincides with the $\psi_2$--loop of \cite{Cooke:2015ila} (or $\tilde{W}_2$ in \cite{Lietti:2017gtc}).
The corresponding conserved supercharges are
\be \label{e34}
\bar\g_i \th^{i\hk}_-, ~~ \ve_{ij}\g^i\th^{j\hk}_+  \qquad\hk = \ho,\hw
\ee
For $\bar\g_i \sim \ve_{ij}\a^j$ they coincide with the ones in (\ref{e31}). This is the degeneracy of WLs discovered in \cite{Cooke:2015ila} and further elaborated in \cite{Griguolo:2015swa, Bianchi:2016vvm, Lietti:2017gtc}.
When $\bar\g_i=0$, $\bar\g_\hi\g^\hi\neq0$, we have a 1/2 BPS WL, $W^\II_{1/2}[0,\bar\g_\hi]$ that preserves supercharges
\be \label{e35}
\bar\g_\hj \th^{i\hj}_-, ~~ \ve_{\hj\hk}\g^\hj\th^{i\hk}_+ \qquad i=1,2
\ee
For $\bar\g_\hi \sim \ve_{\hi\hj} \a^\hj$ they are degenerate with the preserved supercharges of $W^\I_{1/2}[0,\bar\a_\hi]$ defined above.
Finally, when $\bar\g_i\g^i\neq0$ and $\bar\g_\hi\g^\hi\neq0$, we obtain a 1/4 BPS WL, $W^\II_{1/4}[\bar\g_i,\bar\g_\hi]$ with preserved supercharges
\be \label{e36}
\bar\g_i\bar\g_\hj\th^{i\hj}_-, ~~ \ve_{ij}\ve_{\hk\hl}\g^i\g^\hk\th^{j\hl}_+
\ee
All these WLs are cohomologically equivalent to bosonic 1/4 BPS WLs that can be easily determined.

\subsection{M2--/anti--M2--brane duals}\label{secm2}

According to the AdS/CFT correspondence, 1/2 BPS WLs in $U(N)_k \times U(N)_{-k}$ ABJM theory are dual to M2--/anti--M2--branes in $\gabjm$ background \cite{Drukker:2009hy,Lietti:2017gtc}
\be ds^2=R^2 \lt( \frac14 ds^2_{\AdS_4}+ds^2_{ {\rm S}^7/\Z_k} \rt) \ee
More precisely, choosing the $\AdS_4$ metric in the form
\be \label{ads4}
ds^2_{\AdS_4}= u^2 (-dt^2 + dx_1^2 + dx_2^2) + \f{du^2}{u^2}
\ee
and the S$^7$ embedded in $\mathbb{C}^4$ as
\bea \label{s7}
&& ds^2_{\rm{S}^7}=\frac14 \Big[ d\b^2+\cos^2\frac{\b}{2} \big( d\theta_1^2+\sin^2\theta_1 d\varphi_1^2 \big)
                                +\sin^2\frac{\b}{2} \big( d\theta_2^2+\sin^2\theta_2 d\varphi_2^2 \big) \nn\\
&& \phantom{ds^2_{S^7}=}
+\sin^2\frac{\b}{2}\cos^2\frac{\b}{2} (d\chi+\cos\theta_1d\varphi_1-\cos\theta_2d\varphi_2)^2 \nn\\
&& \phantom{ds^2_{S^7}=}
+ \Big( \frac12d\zeta+\cos^2\frac{\b}2\cos\theta_1d\varphi_1+\sin^2\frac{\b}2\cos\theta_2d\varphi_2+\frac12\cos\b d\chi \Big)^2 \Big]
\eea
with the $\mathbb{Z}_k$ identification $\z \sim \z - \f{8\pi}{k}$, the 1/2 BPS operator $W_{1/2}^\I[\bar\a_I]$ reviewed in subsection \ref{ABJM2} is dual to a M2--brane embedded as
$t=\sigma^0, x_1=x_2=0, u=\sigma^1, \zeta=\sigma^2$ and localized at a point specified by the complex vector \cite{Lietti:2017gtc}
\bea\label{unit}
\a^I = (\bar{\a}_I)^\ast &=& (\cos\frac{\b}{2}\cos\frac{\theta_1}{2} \ep^{-\frac{\ii}{4}(2\phi_1+\chi+\zeta)  },
        \cos\frac{\b}{2}\sin\frac{\theta_1}{2} \ep^{-\frac{\ii}{4}(-2\phi_1+\chi+\zeta) }, \nonumber \\
&~~& \sin\frac{\b}{2}\cos\frac{\theta_2}{2} \ep^{-\frac{\ii}{4}(2\phi_2-\chi+\zeta) },
        \sin\frac{\b}{2}\sin\frac{\theta_2}{2} \ep^{-\frac{\ii}{4}(-2\phi_2-\chi+\zeta) })
 \eea
satisfying $\a^I\bar\a_I = 1$. The $\a^I$ vector allows to identify the Killing spinors in $\gabjm$ background corresponding to Poincar\'e and conformal supercharges preserved by the M2--brane solution. One finds that they coincide with supercharges (\ref{preserved}) preserved by $W_{1/2}^\I[\bar\a_I]$.\footnote{For details on this identification we refer the reader to section 4.3 of \cite{Lietti:2017gtc}.}

Similarly, The 1/2 BPS WL $W_{1/2}^\II[\bar\g_I]$ is dual to an anti--M2--brane that wraps a circle specified by a $\bar\g_I$ vector similar to the one in (\ref{unit}). This configuration preserves supercharges in the second line of (\ref{preserved}). When $\bar\a^I = \bar\g^I$ the brane and the anti--brane preserve complementary sets of supercharges \cite{Lietti:2017gtc}, in agreement with the field theory result.

\vskip 7pt
The $\mN=4$ orbifold ABJM theory is dual to M--theory in $\goabjm$ background \cite{Benna:2008zy,Imamura:2008nn,Terashima:2008ba}. Since its 1/2 and 1/4 BPS WLs $W_{1/2}^\I[\bar\a_i,0]$, $W_{1/2}^\I[0,\bar\a_\hi]$ and $W_{1/4}^\I[\bar\a_i,\bar\a_\hi]$ can be obtained from the orbifold decomposition of the 1/2 BPS WL $W_{1/2}^\I[\bar\a_I]$ in ABJM theory, we can easily identify the M2--brane duals of these operators from the dual brane configurations in ABJM theory.

The background can be still described by metrics (\ref{ads4}), (\ref{s7}) with the quotient $\mathbb{Z}_{rk} \times \mathbb{Z}_r$ realized by the identification $\zeta\sim \zeta - \frac{8\pi}{rk}$ , $\chi\sim \chi - \frac{4\pi}{r}$ and $\zeta\sim\zeta - \frac{4\pi}{r}$. Decomposing the R--symmetry indices as in (\ref{vector}) we find that a M2--brane localized at
\bea\label{unit2}
&& \a^i = ( \cos\frac{\b}{2}\cos\frac{\theta_1}{2} \ep^{-\frac{\ii}{4}(2\phi_1+\chi+\zeta)  },
            \cos\frac{\b}{2}\sin\frac{\theta_1}{2} \ep^{-\frac{\ii}{4}(-2\phi_1+\chi+\zeta) }) \nn \\
&&   \a^\hi =( \sin\frac{\b}{2}\sin\frac{\theta_2}{2} \ep^{-\frac{\ii}{4}(-2\phi_2-\chi+\zeta)},
             \sin\frac{\b}{2}\cos\frac{\theta_2}{2} \ep^{-\frac{\ii}{4}(2\phi_2-\chi+\zeta)})
\eea
with $\bar\a_i\a^i+\bar\a_\hi\a^\hi=1$ and generically $\bar\a_i\a^i, \bar\a_\hi\a^\hi \neq 0$, is dual to the 1/4 BPS operator $W_{1/4}^\I[\bar\a_i,\bar\a_\hi]$ introduced in subsection \ref{sec3.2}. The $(\bar\a_i,\bar\a_\hi)$ parameters select the set of supercharges preserved by the corresponding M2--brane configuration \cite{Lietti:2017gtc}, which turn out to coincide with (\ref{e33}). Choosing $\b =0$, that is $\bar\a_\hi=0$, the set of preserved supercharges enhances to (\ref{e31}) and the M2--brane configuration is dual to $W_{1/2}^\I[\bar\a_i,0]$. Similarly, choosing $\b= \pi$ ($\bar\a_i=0$) the set of preserved supercharges coincides with (\ref{e32}) and we obtain the dual configuration of $W_{1/2}^\I[0,\bar\a_\hi]$ .

In a similar fashion, $W_{1/4}^\II[\bar\g_i,\bar\g_\hi]$, $W_{1/2}^\II[\bar\g_i,0]$ and $W_{1/2}^\II[0,\bar\g_\hi]$ operators, arising from the orbifold reduction of WL $W_{1/2}^\II[\bar\g_I]$ in ABJM theory, are dual to anti--M2--branes that wrap circles specified by vectors $(\bar\g_i,\bar\g_\hi)$ of the form (\ref{unit2}).

It is remarkable that the orbifold decomposition provides explicit gravity duals not only for 1/2 BPS operators but also for the 1/4 BPS WLs $W_{1/4}^\I[\bar\a_i,\bar\a_\hi]$ and $W_{1/4}^\II[\bar\g_i,\bar\g_\hi]$, even in the absence of SUSY enhancement.

For generic fermionic 1/6 WLs in ABJM theory the precise gravity duals are not known. Consequently, we are not able to identify precise gravity duals for the generic fermionic 1/4 WLs in $\mN=4$ orbifold ABJM theory obtained by orbifold reduction.

\vskip 10pt

To conclude this section we stress that the orbifold decomposition that leads from $U(rN)_k \times U(rN)_{-k}$ ABJM  to ${\cal N}=4$ orbifold ABJM theory can be generalized to a suitable decomposition of the $U(N)_k \times U(M)_{-k}$ ABJ theory to obtain a $\mN=4$ SCSM theory with circular quiver $\prod_{\ell=1}^r[U(N_{2\ell-1})_k \times U(N_{2\ell})_{-k}]$ \cite{Lietti:2017gtc}. Therefore, the general construction of WLs presented above can be applied also in the more general case of $\mN=4$ SCSM quiver theories. The structure of the superconnections is still the one in (\ref{l2}, \ref{l3}) with definitions (\ref{entries}).
However, since for general circular quiver $\mN=4$ SCSM theories the M--theory dual description is not known, we cannot identify the gravity duals of the corresponding BPS WLs.

\section{BPS WLs in $\mN=4$ SCSM theories: The general approach} \label{secsusy}

Given the previous construction of BPS WLs obtained by direct orbifold decomposition of BPS WLs in ABJM, the natural question arises whether the set of operators in table \ref{tab2} exhausts the whole spectrum of possible 1/4 and 1/2 BPS WLs. To answer this question we now approach the problem by applying a more systematic procedure, which consists in studying directly the SUSY variation of a generic (super)connection $L$ and imposing \cite{Lee:2010hk}
\be \label{e43}
\d L = \p_\t G + \ii [L, G]
\ee
for a suitable supermatrix $G$.

\begin{figure}[htbp]
  \centering
  \includegraphics[width=0.8\textwidth]{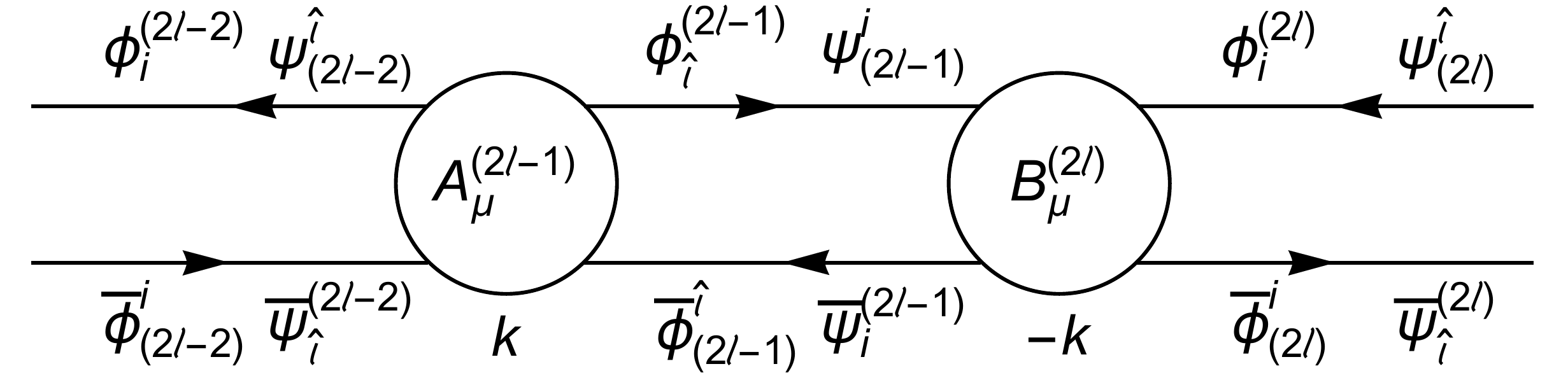}\\
  \caption{A section of quiver diagram of the $\mN=4$ circular quiver SCSM theories with alternating levels. We have $\ell=1,2,\cdots,r$.}\label{qd4}
\end{figure}

 To this end, we consider a $\mN=4$ SCSM theory with gauge group and alternating levels $\prod_{\ell=1}^r[U(N_{2\ell-1})_k \times U(N_{2\ell})_{-k}]$. A section of its quiver is given in fig. \ref{qd4}.

Superconformal transformations are related to Poincar\'e supercharges $P^{i\hi}$, $\bar P_{i\hi}$ and conformal supercharges $S^{i\hi}$, $\bar S_{i\hi}$ as
\be \label{e52}
\d = \ii ( \bar\th^{i\hi} P_{i\hi} + \bar\vth^{i\hi} S_{i\hi} ) = \ii ( \bar P_{i\hi} \th^{i\hi} + \bar S_{i\hi} \vth^{i\hi} )
\ee
The corresponding SUSY transformations on the fields are given in appendix \ref{details} (see eq. (\ref{appA1})).

In what follows we still restrict to WLs defined along the timelike infinite straight line $x^\m=(\t,0,0)$ in Minkowski spacetime.

\subsection{1/4 BPS WLs}

In order to avoid clutter in the presentation we focus on the construction of the particular class of 1/4 WLs preserving Poincar\'e supercharges
\be \label{e44}
\th^{1\ho}_+, ~~ \th^{2\hw}_-
\ee
These correspond to supercharges (\ref{e15}) where we choose
\be
\bar\m_i = (0,1), ~~ \bar\n_\hi = (0,1)
\ee

 First, we observe that for this particular set of parameters, the bosonic 1/4 BPS WL $W^\bos_{1/4}$ in table \ref{tab2} has connection \cite{Ouyang:2015qma,Cooke:2015ila}
\bea \label{e49}
&&  \hspace{-2.2mm}  L_\bos = \diag ( \mA^{(1)}_\bos, \mB^{(2)}_\bos, \cdots, \mA^{(2r-1)}_\bos, \mB^{(2r)}_\bos ) \\
&&  \hspace{-2.2mm}  \mA^{(2\ell-1)}_\bos =  A_0^{(2\ell-1)}
   + \f{2\pi}{k} \Big(- \phi_1^{(2\ell-2)}\bar\phi^1_{(2\ell-2)}
                      + \phi_2^{(2\ell-2)}\bar\phi^2_{(2\ell-2)}
                      - \phi_\ho^{(2\ell-1)}\bar\phi^\ho_{(2\ell-1)}
                      + \phi_\hw^{(2\ell-1)}\bar\phi^\hw_{(2\ell-1)} \Big)  \nn  \\
&&  \hspace{-2.2mm} \mB^{(2\ell)}_\bos =  B_0^{(2\ell)}
   + \f{2\pi}{k} \Big(- \bar\phi^1_{(2\ell)}\phi_1^{(2\ell)}
                      + \bar\phi^2_{(2\ell)}\phi_2^{(2\ell)}
                      - \bar\phi^\ho_{(2\ell-1)}\phi_\ho^{(2\ell-1)}
                      + \bar\phi^\hw_{(2\ell-1)}\phi_\hw^{(2\ell-1)}
                 \Big)   \nn
\eea

To construct fermionic WLs we then consider the most general superconnection $L_\fer$ of the form (\ref{l2}), (\ref{l3}) where
we now allow for the couplings to scalars and fermions in eqs. (\ref{entries}) to depend on the nodes. This choice enlarges the class of BPS WLs obtained by orbifold decomposition and in principle allows for finding more general BPS operators.
Moreover, for computational convenience in (\ref{entries}) we redefine the scalar couplings as
$U_{(2\ell-1)}{}^i_{~j}  = M_{(2\ell-1)}{}^i_{~j} - (\s_3)^i_{~j}$,
$U_{(2\ell-1)}{}^\hi_{~\hj}  = M_{(2\ell-1)}{}^\hi_{~\hj} - (\s_3)^\hi_{~\hj}$,
$U_{(2\ell-1)}{}^i_{~\hj}  = M_{(2\ell-1)}{}^i_{~\hj}$,
$U_{(2\ell-1)}{}^\hi_{~j}  = M_{(2\ell-1)}{}^\hi_{~j}$,
$U_{(2\ell)}{}^i_{~j}  = M_{(2\ell)}{}^i_{~j} - (\s_3)^i_{~j}$,
$U_{(2\ell)}{}^\hi_{~\hj}  = M_{(2\ell)}{}^\hi_{~\hj} - (\s_3)^\hi_{~\hj}$,
$U_{(2\ell)}{}^i_{~\hj}  = M_{(2\ell)}{}^i_{~\hj}$,
$U_{(2\ell)}{}^\hi_{~j}  = M_{(2\ell)}{}^\hi_{~j}$
for all $\ell=1,2,\cdots,r$, with $\s_3$ being the third Pauli matrix.
Given the structure of the quiver in figure \ref{qd4} the superconnection entries then take the form 
\bea \label{e50}
&& \hspace{-2mm} \mA^{(2\ell-1)} =  \mA^{(2\ell-1)}_\bos
   + \f{2\pi}{k} \Big( M_{(2\ell-1)}{}^i_{~j} \phi_i^{(2\ell-2)}\bar\phi^j_{(2\ell-2)}
                      +M_{(2\ell-1)}{}^\hi_{~\hj} \phi_\hi^{(2\ell-1)}\bar\phi^\hj_{(2\ell-1)}
                 \Big) \nn\\
&& \hspace{-2mm} \mB^{(2\ell)} =  \mB^{(2\ell)}_\bos
   + \f{2\pi}{k} \Big( M_{(2\ell)}{}^i_{~j} \bar\phi^j_{(2\ell)}\phi_i^{(2\ell)}
                      +M_{(2\ell)}{}^\hi_{~\hj}\bar\phi^\hj_{(2\ell-1)}\phi_\hi^{(2\ell-1)}
                 \Big) \nn\\
&& \hspace{-2mm} f_1^{(2\ell-1)} = \ii\sqrt{\f{4\pi}{k}} \Big( \bar\a_i^{(2\ell-1)} u_+
                                                - \bar\g_i^{(2\ell-1)} u_- \Big)\psi^i_{(2\ell-1)}, ~~
   f_1^{(2\ell)} = \ii\sqrt{\f{4\pi}{k}} \bar \psi_{\hi}^{(2\ell)} \Big( u_-\b^\hi_{(2\ell)}
                                                                        + u_+ \d^\hi_{(2\ell)} \Big) \nn\\
&& \hspace{-2mm} f_2^{(2\ell-1)} = \ii\sqrt{\f{4\pi}{k}} \bar \psi_{i}^{(2\ell-1)} \Big( u_- \b^i_{(2\ell-1)}
                                                                          + u_+ \d^i_{(2\ell-1)} \Big) , ~~
   f_2^{(2\ell)} = \ii\sqrt{\f{4\pi}{k}} \Big( \bar\a_\hi^{(2\ell)} u_+
                                          - \bar\g_\hi^{(2\ell)} u_- \Big)\psi^\hi_{(2\ell)} \nn\\
&& \hspace{-2mm} h_1^{(2\ell-1)} = \f{2\pi}{k} M_{(2\ell-1)}{}^\hi_{~j} \phi_\hi^{(2\ell-1)}\bar\phi^j_{(2\ell)}, \qquad \qquad \qquad
   h_1^{(2\ell)} = \f{2\pi}{k} M_{(2\ell)}{}^\hi_{~j} \bar\phi^j_{(2\ell)}\phi_\hi^{(2\ell+1)}\nn\\
&& \hspace{-2mm} h_2^{(2\ell-1)} = \f{2\pi}{k} M_{(2\ell-1)}{}^i_{~\hj} \phi_i^{(2\ell)}\bar\phi^\hj_{(2\ell-1)}, \qquad \qquad \qquad
   h_2^{(2\ell)} = \f{2\pi}{k} M_{(2\ell)}{}^i_{~\hj} \bar\phi^\hj_{(2\ell+1)}\phi_i^{(2\ell)}
\eea

To make the fermionic WL BPS, we apply SUSY transformations (\ref{appA1}) to the entries in (\ref{e50}) and impose condition (\ref{e43}) with the choice
\be\label{G}
G = \lt(\ba{cccccccc}
0          & g_1^{(1)} &           &              &              & g_2^{(2r)} \\
g_2^{(1)}  & 0         & g_1^{(2)} &              &              &  \\
           & g_2^{(2)} & \ddots    & \ddots       &              &  \\
           &           & \ddots    & \ddots       & g_1^{(2r-2)} &  \\
           &           &           & g_2^{(2r-2)} & 0            & g_1^{(2r-1)} \\
g_1^{(2r)} &           &           &              & g_2^{(2r-1)} & 0
\ea\rt)
\ee
The detailed structure of the BPS constraints are given in (\ref{appA3}). Solving them for the particular choice (\ref{e44}) of preserved supercharges,  we  obtain
\bea \label{e48}
&& \bar\a_i^{(2\ell-1)} = ( \bar\a_1^{(2\ell-1)} , 0 ), ~~ \qquad
   \bar\a_\hi^{(2\ell)} = ( \bar\a_\ho^{(2\ell)} , 0 ) \nn\\
&& \b^i_{(2\ell-1)}=(\b^1_{(2\ell-1)} , 0), ~~~ \qquad
   \b^\hi_{(2\ell)}=(\b^\ho_{(2\ell)} , 0) \nn\\
&& \bar\g_i^{(2\ell-1)} = ( 0 , \bar\g_2^{(2\ell-1)}), ~~~ \qquad
   \bar\g_\hi^{(2\ell)} = ( 0 , \bar\g_\hw^{(2\ell)}) \nn\\
&& \d^i_{(2\ell-1)}=(0 , \d^2_{(2\ell-1)}), ~~~~ \qquad
   \d^\hi_{(2\ell)}=(0 , \d^\hw_{(2\ell)}) \nn
\eea
\bea
M_{(2\ell-1)}{}^i_{~j} &=&  {\rm diag} \left(
     2\bar\a_\ho^{(2\ell-2)}\b^\ho_{(2\ell-2)}, - 2\bar\g_\hw^{(2\ell-2)}\d^\hw_{(2\ell-2)} \right)  \nn \\
M_{(2\ell-1)}{}^\hi_{~\hj} &=& {\rm diag} \left(
     2\bar\a_1^{(2\ell-1)}\b^1_{(2\ell-1)}, - 2\bar\g_2^{(2\ell-1)}\d^2_{(2\ell-1)} \right) \nn \\
M_{(2\ell)}{}^i_{~j} &=&  {\rm diag} \left(
     2\bar\a_\ho^{(2\ell)}\b^\ho_{(2\ell)} , - 2\bar\g_\hw^{(2\ell)}\d^\hw_{(2\ell)}
     \right) \nn\\
M_{(2\ell)}{}^\hi_{~\hj} &=&  {\rm diag} \left(
     2\bar\a_1^{(2\ell-1)}\b^1_{(2\ell-1)} , - 2\bar\g_2^{(2\ell-1)}\d^2_{(2\ell-1)}
     \right) \nn\\
M_{(2\ell-1)}{}^\hi_{~j} &=&  {\rm diag} \left(
     -2\bar\a_1^{(2\ell-1)}\b^\ho_{(2\ell)} , 2\bar\g_2^{(2\ell-1)}\d^\hw_{(2\ell)}
     \right) \nn\\
M_{(2\ell-1)}{}^i_{~\hj} &=&  {\rm diag} \left(
     -2\bar\a_\ho^{(2\ell)}\b^1_{(2\ell-1)} , 2\bar\g_\hw^{(2\ell)}\d^2_{(2\ell-1)}
     \right) \nn\\
M_{(2\ell)}{}^\hi_{~j} &=&  {\rm diag} \left(
     -2\bar\a_1^{(2\ell+1)}\b^\ho_{(2\ell)} ,  2\bar\g_2^{(2\ell+1)}\d^\hw_{(2\ell)}
     \right) \nn\\
M_{(2\ell)}{}^i_{~\hj} &=& {\rm diag} \left(
     -2\bar\a_\ho^{(2\ell)}\b^1_{(2\ell+1)} , 2\bar\g_\hw^{(2\ell)}\d^2_{(2\ell+1)}
     \right)
\eea
with the remaining parameters subject to
\bea\label{e46}
&& \bar\a_1^{(2\ell-1)} \d^2_{(2\ell-1)} =
   \bar\a_1^{(2\ell-1)} \d^\hw_{(2\ell-2)} =
   \bar\a_1^{(2\ell-1)} \d^\hw_{(2\ell)} = 0 \nn\\
&& \bar\a_\ho^{(2\ell)} \d^2_{(2\ell)} =
   \bar\a_\ho^{(2\ell)} \d^\hw_{(2\ell-1)} =
   \bar\a_\ho^{(2\ell)} \d^\hw_{(2\ell+1)} = 0 \nn\\
&& \bar\g_2^{(2\ell-1)} \b^1_{(2\ell-1)} =
   \bar\g_2^{(2\ell-1)} \b^\ho_{(2\ell-2)} =
   \bar\g_2^{(2\ell-1)} \b^\ho_{(2\ell)} = 0 \nn\\
&& \bar\g_\hw^{(2\ell)} \b^1_{(2\ell)} =
   \bar\g_\hw^{(2\ell)} \b^\ho_{(2\ell-1)} =
   \bar\g_\hw^{(2\ell)} \b^\ho_{(2\ell+1)} = 0
\eea
The most general solution depends on $8r$ parameters
\be
\bar\a_1^{(2\ell-1)}, ~ \bar\a_\ho^{(2\ell)}, ~ \b^1_{(2\ell-1)}, ~ \b^\ho_{(2\ell)}, ~
\bar\g_2^{(2\ell-1)}, ~ \bar\g_\hw^{(2\ell)}, ~ \d^2_{(2\ell-1)}, ~ \d^\hw_{(2\ell)}, ~~
\ell=1,2,\cdots,r
\ee
constrained by (\ref{e46}). The solutions can be classified according to the number of free parameters they depend on,  as we now describe.

\vskip 10pt
\noindent
\underline{Classification of 1/4 BPS WLs}\\

\noindent
1) First, we find four classes of solutions that depend on $4r$ free complex parameters, two parameters for each even node and two for each odd one. They are explicitly given by
\bea \label{e47}
&& \textrm{Class I}: ~~~ \bar\g_2^{(2\ell-1)}=\bar\g_\hw^{(2\ell)}=\d^2_{(2\ell-1)}=\d^\hw_{(2\ell)} = 0 \nn\\
&& \textrm{Class II}: ~~ \bar\a_1^{(2\ell-1)}=\bar\a_\ho^{(2\ell)}=\b^1_{(2\ell-1)}=\b^\ho_{(2\ell)} = 0 \nn\\
&& \textrm{Class III}: ~\b^1_{(2\ell-1)}=\b^\ho_{(2\ell)} =\d^2_{(2\ell-1)}=\d^\hw_{(2\ell)} = 0 \nn\\
&& \textrm{Class IV}: ~ \bar\a_1^{(2\ell-1)}=\bar\a_\ho^{(2\ell)} = \bar\g_2^{(2\ell-1)}=\bar\g_\hw^{(2\ell)}=0
\eea
These solutions are in general non--block--diagonal and generalize the ones in table~\ref{tab2}. In fact, they reduce to the previous ones by imposing that the parameters are the same for every even (odd) node. For example, solutions (\ref{e47}) in Class I are parametrized by $(\bar\a_\ho^{(2\ell)},  \b^\ho_{(2\ell)})$ at even nodes
and $ (\bar\a_1^{(2\ell-1)}, \b^1_{(2\ell-1)})$ at odd nodes. When
\be
\bar\a_1^{(2\ell-1)} = \bar\a_1, ~ ~~ \b^1_{(2\ell-1)} =  \b^1, ~~~ \bar\a_\ho^{(2\ell)} = \bar\a_\ho,   ~~~ \b^\ho_{(2\ell)} =  \b^\ho  ~\qquad
\ell=1,2,\cdots,r
\ee
we obtain solution $W_{1/4}^\I[\bar\m_i,0,0,\bar\n_\hi,p_I]$ in Class I of table~\ref{tab2} corresponding to $\bar\m_i = (0,1)$, $\bar\n_\hi = (0,1)$. \\

\noindent
2) Beyond these four classes, there are additional solutions that depend on a smaller set of parameters.
For example, the solution obtained by setting
\bea
&& \bar\g_2^{(2\ell-1)}=\bar\g_\hw^{(2\ell)} = \d^\hw_{(2\ell)} = 0 ~~ \qquad  \qquad \ell=1,2,\cdots,r \nn\\
&& \bar\a_1^{(1)}=\bar\a_\ho^{(2)}=\bar\a_\ho^{(2r)}=\d^2_{(2\ell-1)}=0 ~~~~~ \ell=2,3,\cdots,r
\eea
depends on $4r-2$ complex parameters and corresponds to a new genuine WL that does not have a counterpart in table \ref{tab2}.

\vskip 10pt

To conclude this section we prove that the most general 1/4 BPS fermionic operator $W^\fer$ corresponding to superconnection $L_\fer$ in (\ref{l2}) or (\ref{l3}) with assignments (\ref{e50}), (\ref{e48}) and (\ref{e46}) is cohomologically equivalent to the bosonic WL (\ref{e49}), that is
\be
\label{cohom2}
W^\fer - W_{1/4}^\bos = {\cal Q} V
\ee
where ${\cal Q}$ is a linear combination of preserved supercharges (\ref{e44}).

In order to prove it  we first split the difference of the connections into a bosonic and a fermionic part
\be
L_\fer - L_\bos \equiv L_B + L_F
\ee
Then, following the prescription in \cite{Drukker:2009hy,Ouyang:2015qma}, we look for a supercharge $\mQ$, one parameter $\k$ and one matrix $\L$ satisfying
\bea
&& \k\L^2 = L_B, ~~ \mQ\L = L_F, ~~ \mQ L_\bos=0 \nn\\
&& \mQ L_F = \p_\t (\ii\k\L) + \ii [L_\bos, \ii\k\L]
\eea
A solution to these equations is given by $ \mQ = \ii ( P^{1\ho}_+ - P^{2\hw}_- )$, $\k=1$ and
\bea\label{lambda}
\L = \lt(\ba{cccccccc}
0           & \l_1^{(1)} &            &               &               & \l_2^{(2r)} \\
\l_2^{(1)}  & 0          & \l_1^{(2)} &               &               &  \\
            & \l_2^{(2)} & \ddots     & \ddots        &               &  \\
            &            & \ddots     & \ddots        & \l_1^{(2r-2)} &  \\
            &            &            & \l_2^{(2r-2)} & 0             & \l_1^{(2r-1)} \\
\l_1^{(2r)} &            &            &               & \l_2^{(2r-1)} & 0
\ea\rt)
\eea
where we have defined
\bea
&& \l_1^{(2\ell-1)} =  \ii \sr{\f{4\pi}{k}} \big( \bar\a_1^{(2\ell-1)}\phi_\ho^{(2\ell-1)}
                                                - \bar\g_2^{(2\ell-1)}\phi_\hw^{(2\ell-1)} \big) \nn\\
&& \l_1^{(2\ell)} =  \ii \sr{\f{4\pi}{k}} \big(  \b^\ho_{(2\ell)}\bar\phi^1_{(2\ell)}
                                               + \d^\hw_{(2\ell)}\bar\phi^2_{(2\ell)} \big) \nn\\
&& \l_2^{(2\ell-1)} =  \ii \sr{\f{4\pi}{k}} \big( - \b^1_{(2\ell-1)}\bar\phi^\ho_{(2\ell-1)}
                                                  - \d^2_{(2\ell-1)}\bar\phi^\hw_{(2\ell-1)} \big) \nn\\
&& \l_2^{(2\ell)} =  \ii \sr{\f{4\pi}{k}} \big( - \bar\a_\ho^{(2\ell)}\phi_1^{(2\ell)}
                                                + \bar\g_\hw^{(2\ell)}\phi_2^{(2\ell)} \big)
\eea

\vskip 10pt
\subsection{1/2 BPS WLs}

We begin by constructing 1/2 BPS WLs that preserve Poincar\'e supercharges
\be \label{e55}
\th^{1\hj}_+, ~~ \th^{2\hj}_-, ~~ \hj = \ho,\hw
\ee
Again, we consider ansatz (\ref{l2}) or (\ref{l3}) for the superconnection with definitions (\ref{e50}) and require the validity of the BPS condition (\ref{e43}) for a suitable choice of the matrix $G$ of the form (\ref{G}). Solving the corresponding constraints we obtain
\bea \label{sol1/2}
&& \bar\a_i^{(2\ell-1)} = ( \bar\a_1^{(2\ell-1)} , 0 ) ,~~
   \b^i_{(2\ell-1)}=(\b^1_{(2\ell-1)} , 0) \nn\\
&& \bar\g_i^{(2\ell-1)} = ( 0 , \bar\g_2^{(2\ell-1)}), ~~
   \d^i_{(2\ell-1)}=(0 , \d^2_{(2\ell-1)})  \\
&& \bar\a_\hi^{(2\ell)} =\b^\hi_{(2\ell)}=\bar\g_\hi^{(2\ell)} =\d^\hi_{(2\ell)}=0 \nn\\
&& \bar\a_1^{(2\ell-1)}\d^2_{(2\ell-1)} = \bar\g_2^{(2\ell-1)}\b^1_{(2\ell-1)} = 0, ~~
\bar\a_1^{(2\ell-1)}\b^1_{(2\ell-1)} + \bar\g_2^{(2\ell-1)} \d^2_{(2\ell-1)} = 1 \nn \\
&& ~~ \nn \\
&& M_{(2\ell-1)}{}^i_{~j} = M_{(2\ell-1)}{}^\hi_{~j} = M_{(2\ell-1)}{}^i_{~\hj} =
   M_{(2\ell)}{}^i_{~j} = M_{(2\ell)}{}^\hi_{~j} = M_{(2\ell)}{}^i_{~\hj} = 0 \nn \\
&& M_{(2\ell-1)}{}^\hi_{~\hj} = {\rm diag} \left(
     2\bar\a_1^{(2\ell-1)}\b^1_{(2\ell-1)} , - 2\bar\g_2^{(2\ell-1)}\d^2_{(2\ell-1)} \right) \nn\\
&& M_{(2\ell)}{}^\hi_{~\hj} =  {\rm diag} \left(
     2\bar\a_1^{(2\ell-1)}\b^1_{(2\ell-1)} , - 2\bar\g_2^{(2\ell-1)}\d^2_{(2\ell-1)}
     \right) \nn
\eea
Notably, all the solutions correspond to block--diagonal superconnections $L_{1/2}$ with $r$ independent blocks of the form
\be
L_{1/2} = \lt(\ba{cccccccc}
\mA^{(1)} & f_1^{(1)} &        &        &              & \\
f_2^{(1)} & \mB^{(2)} &        &        &              & \\
          &           & \ddots & \ddots &              & \\
          &           & \ddots & \ddots &              & \\
          &           &        &        & \mA^{(2r-1)} & f_1^{(2r-1)} \\
          &           &        &        & f_2^{(2r-1)} & \mB^{(2r)}
\ea\rt)
\ee
In each block we have two possible choices of the parameters
\bea
&& \textrm{1st choice}: ~~~ \bar\a_1^{(2\ell-1)}\b^1_{(2\ell-1)} = 1, ~~ \bar\g_2^{(2\ell-1)}=\d^2_{(2\ell-1)}=0  \nn\\
&& \textrm{2nd choice}: ~~ \bar\a_1^{(2\ell-1)} = \b^1_{(2\ell-1)} = 0, ~~ \bar\g_2^{(2\ell-1)}\d^2_{(2\ell-1)}=1
\eea
Therefore, there are in total $2^r$ different 1/2 BPS WLs, each depending on $r$ free complex parameters. The 1/2 BPS WLs $W_{1/2}^\I[\bar\a_i,0]$ and $W_{1/2}^\II[\bar\g_i,0]$ in section~\ref{sec3.2}, with respectively $\bar\a_i=(\bar\a_1,0)$ and $\bar\g_i=(0,\bar\g_2)$, are just special cases of these general 1/2 BPS operators.

Similarly, we can construct operators that preserve Poincar\'e supercharges
\be \label{e56}
\th^{i\ho}_+, ~~ \th^{i\hw}_-, ~~ i = 1,2
\ee
This time the couplings entering superconnection (\ref{l2}) or (\ref{l3}) with definitions (\ref{e50}) are fixed by the following set of constraints
\bea
&& \bar\a_\hi^{(2\ell)} = ( \bar\a_\ho^{(2\ell)} , 0 ) ,~~
   \b^\hi_{(2\ell)}=(\b^\ho_{(2\ell)} , 0) \\
&& \bar\g_\hi^{(2\ell-1)} = ( 0 , \bar\g_\hw^{(2\ell)}), ~~
   \d^\hi_{(2\ell-1)}=(0 , \d^\hw_{(2\ell)})  \nn\\
&& \bar\a_i^{(2\ell-1)} =\b^i_{(2\ell-1)}=\bar\g_i^{(2\ell-1)} =\d^i_{(2\ell-1)}=0 \nn\\
&& \bar\a_\ho^{(2\ell)}\d^\hw_{(2\ell)} = \bar\g_\hw^{(2\ell)}\b^1_{(\hw\ell)} = 0, ~~
\bar\a_\ho^{(2\ell)}\b^\ho_{(2\ell)} + \bar\g_\hw^{(2\ell)} \d^2_{(\hw\ell)} = 1 \nn \\
&& ~~ \nn \\
&& M_{(2\ell-1)}{}^\hi_{~\hj} = M_{(2\ell-1)}{}^\hi_{~j} = M_{(2\ell-1)}{}^i_{~\hj} =
   M_{(2\ell)}{}^\hi_{~\hj} = M_{(2\ell)}{}^\hi_{~j} = M_{(2\ell)}{}^i_{~\hj} = 0 \nn \\
&& M_{(2\ell-1)}{}^i_{~j} =  {\rm diag} \left(
     2\bar\a_\ho^{(2\ell-2)}\b^\ho_{(2\ell-2)} , - 2\bar\g_\hw^{(2\ell-2)}\d^\hw_{(2\ell-2)}
     \right) \nn\\
&& M_{(2\ell)}{}^i_{~j} = {\rm diag} \left(
     2\bar\a_\ho^{(2\ell)}\b^\ho_{(2\ell)} , - 2\bar\g_\hw^{(2\ell)}\d^\hw_{(2\ell)}
     \right) \nn
\eea
The corresponding superconnection is also block--diagonal with $r$ independent blocks
\be
\tilde{L}_{1/2} = \lt(\ba{cccccccc}
\mA^{(1)}  &           &           &        &        & f_2^{(2r)} \\
           & \mB^{(2)} & f_1^{(2)} &        &        & \\
           & f_2^{(2)} & \mA^{(3)} &        &        & \\
           &           &           & \ddots & \ddots & \\
           &           &           & \ddots & \ddots & \\
f_1^{(2r)} &           &           &        &        & \mB^{(2r)}
\ea\rt)
\ee
and in each block the parameters can be chosen in two different ways
\bea
&& \textrm{1st choice}: ~~~\bar\a_\ho^{(2\ell)}\b^\ho_{(2\ell)} = 1, ~~ \bar\g_\hw^{(2\ell)}=\d^\hw_{(2\ell)}=0  \nn\\
&& \textrm{2nd choice}: ~~ \bar\a_\ho^{(2\ell)} = \b^\ho_{(2\ell)} = 0, ~~ \bar\g_\hw^{(2\ell)}\d^\hw_{(2\ell)}=1
\eea
Therefore, there are still $2^r$ different operators, each one depending on $r$ free complex parameters. The 1/2 BPS WLs $W_{1/2}^\I[0,\bar\a_\hi]$ with $\bar\a_\hi=(\bar\a_\ho,0)$ and $W_{1/2}^\II[0,\bar\g_\hi]$ with $\bar\g_\hi=(0,\bar\g_\hw)$ in section~\ref{sec3.2} are special cases of this general class of 1/2 BPS WLs.

The 1/2 BPS WLs in this section correspond to superconnections that are simply direct sums of the double--node 1/2 BPS superconnections already introduced in \cite{Ouyang:2015qma,Cooke:2015ila,Ouyang:2015hta}. There are in fact no new 1/2 BPS solutions.

Note that supercharges (\ref{e44}) are included in both (\ref{e55}) and (\ref{e56}), so the 1/2 BPS WLs just found are special cases of the  fermionic 1/4 BPS WLs constructed in the previous subsection, as one can easily check.

\section{Summary and discussion}\label{seccon}

The main result of this paper is the construction of new 1/4 BPS WLs in $\mN=4$ circular quiver SCSM theories with gauge group and levels $\prod_{\ell=1}^r [ U(N_{2\ell-1})_k \times U(N_{2\ell})_{-k} ]$.

First of all, we have considered $[U(N)_k\times U(N)_{-k}]^r$  $\mN=4$   orbifold ABJM models.  By performing the orbifold decomposition of 1/2 and 1/6 BPS WLs in ABJ(M) we have not only recovered the already known WLs, but also found new WLs described by more general superconnections that are not block--diagonal like the ones considered so far in the literature \cite{Ouyang:2015qma,Cooke:2015ila,Ouyang:2015hta,Ouyang:2015bmy,Lietti:2017gtc}. As a consequence, the structure of the corresponding WLs is more general and not reducible to ABJ(M)--like double--node operators. These new WLs are 1/4 BPS and get enhanced to the already known 1/2 BPS for special values of the matter couplings appearing in the superconnection. They can be classified into four classes, and each class is parametrized by four free complex numbers. Setting all the parameters to zero we obtain the bosonic 1/4 BPS WL. Additionally, we  have  identified the corresponding M2--/anti--M2--brane duals. The novelty here is that we find the explicit M2 configuration dual to 1/4 BPS operators $W^\I_{1/4}[\bar\a_i,\bar\a_\hi]$ and  $W^\II_{1/4}[\bar\g_i,\bar\g_\hi]$, beyond the already known duals of 1/2 BPS WLs \cite{Lietti:2017gtc}.

These findings have been confirmed and generalized by using a more systematic approach that consists in studying the SUSY transformations of the most general superconnection in a $\mN=4$ SCSM theory and imposing its invariance under a given subset of supercharges. We have obtained the  complete spectrum of 1/4 and 1/2 BPS WLs that contains a generalization of the four classes already mentioned plus some extra operators that fall outside these classes. The complete spectrum of fermionic 1/4 and 1/2 BPS WLs in $\mN=4$ circular quiver SCSM theories can be summarized as follows.

\vskip 7pt
\noindent
\underline{1/4 BPS WLs}: With the set of preserved supercharges being fixed, the most general 1/4 BPS WL is the holonomy of a superconnection (\ref{l2}) or (\ref{l3}) that depends on $8r$ complex parameters subject to $12r$ non--linear constraints. There are four classes of solutions to these constraints, and in each class the WL has $4r$ free complex parameters. There are also other solutions that do not fall into the four classes, and such WLs necessarily have fewer free parameters.

\noindent
\underline{1/2 BPS WLs}: 1/2 BPS WLs always correspond to block diagonal superconnections with $r$ blocks. Fixing the set of preserved supercharges, there are $2^r$ different choices of the parameters, and for each choice the operator depends on $r$ free complex parameters.

\vskip 7pt

Our investigation of the BPS nature of WLs has been carried out on operators defined on timelike straight lines in Minkowski spacetime. Performing a Wick rotation followed by a conformal transformation we can map them to WLs on circular contours in Euclidean space. For completeness, we report  in appendix~\ref{appcir} the explicit forms of the circular 1/4 BPS WLs, which are particularly relevant for other purposes. Indeed, it is the expectation value of circular bosonic 1/4 BPS WL $W_{1/4}^\bos[\bar\m_i,0,0,\bar\n_\hi] $ in Euclidean space that can be computed exactly using localization techniques \cite{Kapustin:2009kz}.   In fact, given the circular bosonic 1/4 BPS WL
\be
W_{1/4}^{\bos} = \Tr\mP\ep^{-\ii\oint d\t L_{\bos}}
\ee
where $L_\bos$ is of the form (\ref{LbosE}), being the theory invariant under $SU(2)_L \times SU(2)_R$ R--symmetry, the expectation value is independent of the particular choice of the $\bar{\mu}_i, \bar{\nu}_{\hi}$ parameters.Since it is block diagonal with blocks $L_{\bos}^{(a)}$, $a=1,2,\cdots,2r$, the expectation value reduces to a sum over the blocks
\be \label{MM}
\lag W_{1/4}^{\bos} \rag_1 = \sum_{a=1}^{2r} \lag W_{1/4}^{\bos,{(a)}} \rag_1
\ee
 The subscript "1" indicates that the result is at framing one, the regularization scheme automatically selected by the localization procedure \cite{Kapustin:2009kz}.   The weak coupling expansion of the $(a)$--block reads \cite{Griguolo:2015swa,Bianchi:2016vvm}
\be\label{weakexp}
\lag W_{1/4}^{\bos,{(a)}} \rag_1 = N_a \Big[ 1 + \f{(-)^{a+1}\ii\pi}{k}N_a
+ \f{\pi^2}{6k^2} ( -4N_a^2 + 3N_a N_{a-1}+ 3N_a N_{a+1} +1 ) + O\Big(\f1{k^3}\Big) \Big]
\ee
This immediately leads to an interesting feature of the full family of WLs that we have introduced.
All the fermionic 1/4 and 1/2 BPS WLs are in the same ${\cal Q}$--cohomological class of $W_{1/4}^\bos[\bar\m_i,0,0,\bar\n_\hi] $, where ${\cal Q}$ is the charge used to localize the path integral. Therefore, if the cohomological equivalence in not broken by quantum anomalies, all the fermionic WLs should have the same framing one expectation value (\ref{MM}), (\ref{weakexp}).

The weak coupling expansion (\ref{weakexp}) has to match the result obtained from perturbation theory. Since the perturbative result is at trivial framing, this requires to correctly identify and remove the framing factor from the matrix model result. In the ABJ(M) case, for the bosonic 1/6 BPS WL this has been discussed in \cite{Bianchi:2016yzj,Bianchi:2016gpg} where it has been proved that both gauge and matter sectors contribute to build up the correct framing phase.

In the case of $\mN=4$ circular quiver SCSM theories, generalizing the results of \cite{Bianchi:2013rma,Griguolo:2013sma} we can compute the two--loop expectation values of all 1/2 and 1/4 BPS WLs. We expect the framing--zero results to depend non--trivially on the parameters \cite{work}. Since the matrix model prediction (\ref{MM}), (\ref{weakexp}) is parameter independent, we conclude that a perturbative calculation done at framing one should enlighten a non--trivial conspiracy between gauge and matter sectors, whose parameter dependent contributions should cancel each other. We note that a similar pattern should arise also for parametric fermionic 1/6 BPS WLs in ABJ(M) theory listed in table \ref{tab1}. We will report a detailed study of this issue in \cite{work}.

In $\mN=4$ circular quiver SCSM theories a further related subtlety arises concerning degenerate WLs, that is fermionic 1/2  BPS operators in Classes I and II that preserve the same set of supercharges (see appendix \ref{overlapping}).
Although at quantum level they are expected to have the same framing--one expectation value, being both ${\cal Q}$--equivalent to the bosonic 1/4 BPS WL, it has been proved that at framing--zero they start being different at three loops \cite{Griguolo:2015swa,Bianchi:2016vvm}, at least for quiver theories with different group ranks.
More generally, an all--loop argument proves that they are necessarily different at any odd order, unless vanishing. This potential tension between the matrix model prediction and the perturbative calculation is presently an open question that definitively requires further investigation.

Finally, BPS WLs have been shown to be related to physical quantities like the energy radiated by a heavy quark slowly moving in a gauge background (Bremsstrahlung function) and the cusp anomalous dimension that governs the UV divergent behavior of WLs close to a cusp. In ABJM theory an exact prescription has been proposed \cite{Lewkowycz:2013laa, Bianchi:2014laa} that gives the Bremsstrahlung function for 1/2 BPS quark configurations in terms of the framing phase of the 1/6 BPS bosonic WL.  
The cusp anomalous dimension and the Bremsstrahlung function are in principle amenable of exact computation via integrability, along the lines of the $\mN=4 $ SYM case \cite{Correa:2012hh,Drukker:2012de}. This would require using the conjectural exact form of the interpolating $h(\l)$ function \cite{Gromov:2014eha}. Therefore, BPS WLs turn out to be a potentially crucial tool to test the exact $h(\lambda)$ and, more generally, the integrability underlying the AdS/CFT correspondence (see \cite{Cavaglia:2014exa,Bombardelli:2017vhk} for some preliminary results). It would be interesting to compute the cusp anomalous dimension for a cusp with generic fermionic 1/4 BPS operators on the two rays and study its dependence on the parameters in $\mN=4 $ SCSM models.  This would also open the possibility to define and compute the corresponding Bremsstrahlung function along the lines of \cite{Lewkowycz:2013laa} and \cite{Bianchi:2014laa, Bianchi:2017svd}.

\acknowledgments

We thank Jun-Bao Wu for helpful discussions.
This work has been supported in part by Italian Ministero dell'Istruzione, Universit\`a e Ricerca (MIUR) and Istituto
Nazionale di Fisica Nucleare (INFN) through the ``Gauge Theories, Strings, Supergravity'' (GSS) research project.
J.-j.Z. is supported by the ERC Starting Grant 637844-HBQFTNCER.

\appendix

\section{Technical details} \label{details}

Here we collect some technical details necessary to follow the general discussion of BPS WLs in section \ref{secsusy}.

Given a generic ${\cal N} = 4$ SCSM theory with gauge group $\prod_{\ell=1}^r [ U(N_{2\ell-1})_k \times U(N_{2\ell})_{-k} ]$,
using the field labeling of figure~\ref{qd4} the SUSY transformations read explicitly
{\small \bea \label{appA1}
&&  \hspace{-3mm}
  \d A_\m^{(2\ell-1)}=-\f{2\pi}{k} \Big[ \Big( \phi_i^{(2\ell-2)}\bar\psi_\hi^{(2\ell-2)}
                                             - \phi_\hi^{(2\ell-1)}\bar\psi_i^{(2\ell-1)} \Big) \g_\m \e^{i\hi}
                   +\bar\e_{i\hi}\g_\m \Big( \psi^\hi_{(2\ell-2)}\bar\phi^i_{(2\ell-2)}
                                            - \psi^i_{(2\ell-1)}\bar\phi^\hi_{(2\ell-1)} \Big) \Big] \nn\\
&&  \hspace{-3mm}
  \d B_\m^{(2\ell)}=-\f{2\pi}{k} \Big[ \Big( \bar\psi_\hi^{(2\ell)}\phi_i^{(2\ell)}
                                           - \bar\psi_i^{(2\ell-1)}\phi_\hi^{(2\ell-1)} \Big) \g_\m\e^{i\hi}
                   +\bar\e_{i\hi}\g_\m \Big( \bar\phi^i_{(2\ell)}\psi^\hi_{(2\ell)}
                                            - \bar\phi^\hi_{(2\ell-1)}\psi^i_{(2\ell-1)} \Big) \Big] \nn\\
&&  \hspace{-3mm}
\d\phi_\hi^{(2\ell-1)} = -\ii\bar\e_{i\hi}\psi^i_{(2\ell-1)}, ~~
\d\bar\phi^\hi_{(2\ell-1)} = -\ii\bar\psi_i^{(2\ell-1)}\e^{i\hi} , ~~
\d\phi_i^{(2\ell)} = \ii\bar\e_{i\hi}\psi^\hi_{(2\ell)}, ~~
\d\bar\phi^i_{(2\ell)} = \ii\bar\psi_\hi^{(2\ell)}\e^{i\hi} \nn\\
&& \hspace{-3mm}
   \d\psi^i_{(2\ell-1)}= \g^\m\e^{i\hi} D_\m\phi_\hi^{(2\ell-1)} +  \vth^{i\hi} \phi_\hi^{(2\ell-1)}
                      -\f{4\pi}{k}\e^{j\hj} \Big(  \phi_\hj^{(2\ell-1)}\bar\phi^i_{(2\ell)}\phi_j^{(2\ell)}
                                                  -\phi_j^{(2\ell-2)}\bar\phi^i_{(2\ell-2)}\phi_\hj^{(2\ell-1)} \Big) \nn\\
&& \hspace{-3mm}\phantom{\d\psi^i_{(2\ell-1)}=}
                      +\f{2\pi}{k}\e^{i\hi} \Big( \phi_\hi^{(2\ell-1)}\bar\phi^\hj_{(2\ell-1)}\phi_\hj^{(2\ell-1)}
                                                 +\phi_\hi^{(2\ell-1)}\bar\phi^j_{(2\ell)}\phi_j^{(2\ell)} \nn\\
&& \hspace{-3mm}\phantom{\d\psi^i_{(2\ell-1)}=}
                                                 -\phi_\hj^{(2\ell-1)}\bar\phi^\hj_{(2\ell-1)}\phi_\hi^{(2\ell-1)}
                                                 -\phi_j^{(2\ell-2)}\bar\phi^j_{(2\ell-2)}\phi_\hi^{(2\ell-1)}      \Big) \nn\\
&& \hspace{-3mm}
   \d\bar\psi_i^{(2\ell-1)}= -\bar\e_{i\hi}\g^\m D_\m\bar\phi^\hi_{(2\ell-1)} + \bar\vth_{i\hi} \bar\phi^\hi_{(2\ell-1)}
                      +\f{4\pi}{k}\bar\e_{j\hj} \Big(  \bar\phi^\hj_{(2\ell-1)}\phi_i^{(2\ell-2)}\bar\phi^j_{(2\ell-2)}
                                                      -\bar\phi^j_{(2\ell)}\phi_i^{(2\ell)}\bar\phi^\hj_{(2\ell-1)} \Big) \nn\\
&& \hspace{-3mm}\phantom{\d\bar\psi_i^{(2\ell-1)}=}
                      -\f{2\pi}{k}\bar\e_{i\hi} \Big(  \bar\phi^\hi_{(2\ell-1)}\phi_\hj^{(2\ell-1)}\bar\phi^\hj_{(2\ell-1)}
                                                      +\bar\phi^\hi_{(2\ell-1)}\phi_j^{(2\ell-2)}\bar\phi^j_{(2\ell-2)} \nn\\
&& \hspace{-3mm}\phantom{\d\bar\psi_i^{(2\ell-1)}=}
                                                      -\bar\phi^\hj_{(2\ell-1)}\phi_\hj^{(2\ell-1)}\bar\phi^\hi_{(2\ell-1)}
                                                      -\bar\phi^j_{(2\ell)}\phi_j^{(2\ell)}\bar\phi^\hi_{(2\ell-1)} \big) \nn\\
&& \hspace{-3mm}
   \d\psi^\hi_{(2\ell)}=-\g^\m\e^{i\hi} D_\m\phi_i^{(2\ell)} - \vth^{i\hi} \phi_i^{(2\ell)}
                      -\f{4\pi}{k}\e^{j\hj} \Big(  \phi_\hj^{(2\ell+1)}\bar\phi^\hi_{(2\ell+1)}\phi_j^{(2\ell)}
                                                  -\phi_j^{(2\ell)}\bar\phi^\hi_{(2\ell-1)}\phi_\hj^{(2\ell-1)} \Big) \nn\\
&& \hspace{-3mm}\phantom{\d\psi^\hi_{(2\ell)}=}
                      -\f{2\pi}{k}\e^{i\hi} \Big(  \phi_i^{(2\ell)}\bar\phi^\hj_{(2\ell-1)}\phi_\hj^{(2\ell-1)}
                                                  +\phi_i^{(2\ell)}\bar\phi^j_{(2\ell)}\phi_j^{(2\ell)}  
                                                  -\phi_\hj^{(2\ell+1)}\bar\phi^\hj_{(2\ell+1)}\phi_i^{(2\ell)}
                                                  -\phi_j^{(2\ell)}\bar\phi^j_{(2\ell)}\phi_i^{(2\ell)}  \Big) \nn\\
&& \hspace{-3mm}
   \d\bar\psi_\hi^{(2\ell)} = \bar\e_{i\hi}\g^\m D_\m\bar\phi^i_{(2\ell)} - \bar\vth_{i\hi}\bar\phi^i_{(2\ell)}
                      +\f{4\pi}{k}\bar\e_{j\hj} \Big(  \bar\phi^\hj_{(2\ell-1)}\phi_\hi^{(2\ell-1)}\bar\phi^j_{(2\ell)}
                                                      -\bar\phi^j_{(2\ell)}\phi_\hi^{(2\ell+1)}\bar\phi^\hj_{(2\ell+1)} \Big) \\
&& \hspace{-3mm}\phantom{\d\psi_\hi^{(2\ell)}=}
                      +\f{2\pi}{k}\bar\e_{i\hi} \Big(  \bar\phi^i_{(2\ell)}\phi_\hj^{(2\ell+1)}\bar\phi^\hj_{(2\ell+1)}
                                                       +\bar\phi^i_{(2\ell)}\phi_j^{(2\ell)}\bar\phi^j_{(2\ell)}  
                                                       -\bar\phi^\hj_{(2\ell-1)}\phi_\hj^{(2\ell-1)}\bar\phi^i_{(2\ell)}
                                                       -\bar\phi^j_{(2\ell)}\phi_j^{(2\ell)}\bar\phi^i_{(2\ell)} \Big) \nn
\eea}
where the SUSY parameters are
\be
\e^{i\hi}=\th^{i\hi}+x^\m\g_\m\vth^{i\hi}, \qquad \qquad \bar\e_{i\hi}=\bar\th_{i\hi}-\bar\vth_{i\hi}x^\m\g_\m
\ee
We apply these transformations to the most general superconnection of the form (\ref{l2}), (\ref{l3}) and impose that the result can be written as in
(\ref{e43}) with a matrix $G$ given in (\ref{G}). This condition translates into a set of constraints on the superconnections entries (\ref{entries}) that read
\bea\label{appA3}\label{appA3}
&& \hspace{-6mm}
   \d \mA^{(2\ell-1)} = \ii ( f_1^{(2\ell-1)} g_2^{(2\ell-1)} + f_2^{(2\ell-2)} g_1^{(2\ell-2)}
                            - g_1^{(2\ell-1)} f_2^{(2\ell-1)} - g_2^{(2\ell-2)} f_1^{(2\ell-2)} ) \nn\\
&& \hspace{-6mm}
   \d \mB^{(2\ell-1)} = \ii ( f_1^{(2\ell)} g_2^{(2\ell)} + f_2^{(2\ell-1)} g_1^{(2\ell-1)}
                            - g_1^{(2\ell)} f_2^{(2\ell)} - g_2^{(2\ell-1)} f_1^{(2\ell-1)} ) \nn\\
&& \hspace{-6mm}
   \d h_1^{(2\ell-1)} = \ii ( f_1^{(2\ell-1)} g_1^{(2\ell)} - g_1^{(2\ell-1)} f_1^{(2\ell)} ), ~~
   \d h_1^{(2\ell)} = \ii ( f_1^{(2\ell)} g_1^{(2\ell+1)} - g_1^{(2\ell)} f_1^{(2\ell+1)} ) \nn\\
&& \hspace{-6mm}
   \d h_2^{(2\ell-1)} = \ii ( f_2^{(2\ell)} g_2^{(2\ell-1)} - g_2^{(2\ell)} f_2^{(2\ell-1)} ), ~~
   \d h_2^{(2\ell)} = \ii ( f_2^{(2\ell+1)} g_2^{(2\ell)} - g_2^{(2\ell+1)} f_2^{(2\ell)} ) \nn\\
&& \hspace{-6mm}
   \d f_1^{(2\ell-1)} = \p_\t g_1^{(2\ell-1)} + \ii ( \mA^{(2\ell-1)}g_1^{(2\ell-1)} - g_1^{(2\ell-1)} \mB^{(2\ell)}
                                                    + h_1^{(2\ell-1)}g_2^{(2\ell)} - g_2^{(2\ell-2)}h_1^{(2\ell-2)} ) \nn\\
&& \hspace{-6mm}
   \d f_1^{(2\ell)} = \p_\t g_1^{(2\ell)} + \ii ( \mB^{(2\ell)}g_1^{(2\ell)} - g_1^{(2\ell)} \mA^{(2\ell+1)}
                                                + h_1^{(2\ell)}g_2^{(2\ell+1)} - g_2^{(2\ell-1)}h_1^{(2\ell-1)} ) \nn \\
&& \hspace{-6mm}
   \d f_2^{(2\ell-1)} = \p_\t g_2^{(2\ell-1)} + \ii ( \mB^{(2\ell)}g_2^{(2\ell-1)} - g_2^{(2\ell-1)} \mA^{(2\ell-1)}
                                                    + h_2^{(2\ell-2)}g_1^{(2\ell-2)} - g_1^{(2\ell)}h_2^{(2\ell-1)} ) \nn\\
&& \hspace{-6mm}
   \d f_2^{(2\ell)} = \p_\t g_2^{(2\ell)} + \ii ( \mA^{(2\ell+1)}g_2^{(2\ell)} - g_2^{(2\ell)} \mB^{(2\ell)}
+ h_2^{(2\ell-1)}g_1^{(2\ell-1)} - g_1^{(2\ell+1)}h_2^{(2\ell)} ) \nn\\
&& ~~ \nn \\
&& \hspace{-6mm}
   h_1^{(2\ell-1)}g_1^{(2\ell+1)} - g_1^{(2\ell-1)}h_1^{(2\ell)} =
   h_2^{(2\ell-1)}g_2^{(2\ell-2)} - g_2^{(2\ell)}h_2^{(2\ell-2)} = 0  \nn\\
&& \hspace{-6mm}
   h_1^{(2\ell-2)}g_1^{(2\ell)} - g_1^{(2\ell-2)}h_1^{(2\ell-1)} =
   h_2^{(2\ell)}g_2^{(2\ell-1)} - g_2^{(2\ell+1)}h_2^{(2\ell-1)} = 0
\eea
These constraints allow for non--trivial solutions depending on the number of preserved SUSY charges, as discussed in section \ref{secsusy} (see eqs. (\ref{e48}), (\ref{e46}) and (\ref{sol1/2})).

\section{Supercharge overlapping of 1/2 BPS WLs}\label{secsup}

This appendix is a supplement to \cite{Lietti:2017gtc}, and we discuss the amount of overlapping supercharges for general 1/2 BPS WLs both in ABJ(M) theory and general $\mN=4$ circular quiver SCSM theories with alternating levels.

\subsection{ABJ(M) theory}

Given the two classes of 1/2 BPS WLs $W^\I_{1/2}[\bar\a_I]$, $W^\II_{1/2}[\bar\g_I]$ reviewed in subsection \ref{ABJM2}, we can distinguish three different cases.
\begin{itemize}

  \item $W^\I_{1/2}[\bar\a_I]$ and $W^\I_{1/2}[\bar\a_I']$. When $\bar\a_I' \pp \bar\a_I$ they are in fact the same WL and trivially preserve the same supercharges. Otherwise, when $\bar\a_I' \sp \bar\a_I$, they share 1/3 of the preserved supercharges
      \be \bar\a_I\bar\a_J'\th^{IJ}_+, ~~ \ve_{IJKL}\a^I\a'^{J}\th^{KL}_- \ee

  \item $W^\I_{1/2}[\bar\a_I]$ and $W^\II_{1/2}[\bar\g_I]$. When $\bar\a_I\g^I=0$, they share 2/3 of the preserved supercharges
      \be \bar\a_I R^J_{~K}\th^{IK}_+ , ~~ \bar\g_I R^J_{~K}\th^{IK}_- \ee
      with $R^J_{~K} = \d^J_K - \f{\a^J\bar\a_K}{|\a|^2} - \f{\g^J\bar\g_K}{|\g|^2}$. Otherwise, when $\bar\a_I\g^I \neq 0$, there is no overlapping of preserved supercharges. In particular, for $\bar\a_I = \bar\g_I$ they preserve complementary sets of supercharges.

  \item $W^\II_{1/2}[\bar\g_I]$ and $W^\II_{1/2}[\bar\g_I']$. When $\bar\g_I' \pp \bar\g_I$, they trivially preserve the same supercharges. Otherwise, when $\bar\g_I' \sp \bar\g_I$, they share 1/3 of the preserved supercharges
      \be \bar\g_I\bar\g_J'\th^{IJ}_-, ~~ \ve_{IJKL}\g^I\g'^{J}\th^{KL}_+ \ee

\end{itemize}
All these results are consistent with the examples discussed in \cite{Lietti:2017gtc}. In particular, in ABJ(M) theory, there are no non--trivial cases of different 1/2 BPS WLs preserving the same set of supercharges.

\subsection{$\mN=4$ circular quiver SCSM theories with alternating levels}\label{overlapping}

As discussed in the main text, in this case we have four classes of 1/2 BPS WLs, $W^\I_{1/2}[\bar\a_i,0]$, $W^\I_{1/2}[0,\bar\a_\hi]$, $W^\II_{1/2}[\bar\g_i,0]$ and $W^\II_{1/2}[0,\bar\g_\hi]$. We need then to distinguish ten different cases.
\begin{itemize}

  \item $W^\I_{1/2}[\bar\a_i,0]$ and $W^\I_{1/2}[\bar\a_i',0]$. When $\bar\a_i' \pp \bar\a_i$ they are basically the same operator and trivially preserve the same supercharges. Otherwise, when $\bar\a_i' \sp \bar\a_i$ they do not share any preserved supercharges.

  \item $W^\I_{1/2}[\bar\a_i,0]$ and $W^\I_{1/2}[0,\bar\a_\hi']$. They share 1/2 of the preserved supercharges
      \be \bar\a_i\bar\a'_\hj\th^{i\hj}_+, ~~ \ve_{ij}\ve_{\hk\hl}\a^i\a'^\hk\th^{j\hl}_- \ee

  \item $W^\I_{1/2}[\bar\a_i,0]$ and $W^\II_{1/2}[\bar\g_i,0]$. When $\bar\a_i\g^i=0$, they preserve the same supercharges
      \be \bar\a_i \th^{i\hj}_+ \sim \ve_{ik}\g^k\th^{i\hj}_+ , ~~
          \bar\g_i \th^{i\hj}_- \sim \ve_{ik}\a^k\th^{i\hj}_- \ee
      Otherwise, when $\bar\a_i\g^i \neq 0$, they do not share any preserved supercharge.

  \item $W^\I_{1/2}[\bar\a_i,0]$ and $W^\II_{1/2}[0,\bar\g_\hi]$. They share 1/2 of the preserved supercharges
      \be \ve_{\hi\hj}\bar\a_k \g^\hi\th^{k\hj}_+, ~~
          \ve_{ij}\a^i\bar\g_\hk\th^{j\hk}_- \ee

  \item $W^\I_{1/2}[0,\bar\a_\hi]$ and $W^\I_{1/2}[0,\bar\a_\hi']$. When $\bar\a_\hi' \pp \bar\a_\hi$, they trivially preserve the same supercharges. Otherwise, they do not share any preserved supercharge.

  \item $W^\I_{1/2}[0,\bar\a_\hi]$ and $W^\II_{1/2}[\bar\g_i,0]$. They share 1/2 of the preserved supercharges
      \be \ve_{ij}\bar\a_\hk\g^i\th^{j\hk}_+, ~~ \ve_{\hi\hj}\a^\hi\bar\g_k\th^{k\hj}_- \ee

  \item $W^\I_{1/2}[0,\bar\a_\hi]$ and $W^\II_{1/2}[0,\bar\g_\hi]$. When $\bar\a_\hi\g^\hi=0$, they preserve the same supercharges
      \be \bar\a_\hj \th^{i\hj}_+ \sim \ve_{\hj\hk}\g^\hk\th^{i\hj}_+ , ~~
          \bar\g_\hj \th^{i\hj}_- \sim \ve_{\hj\hk}\a^\hk\th^{i\hj}_- \ee
      Otherwise, when $\bar\a_\hi\g^\hi \neq 0$, they do not share any preserved supercharge.

  \item $W^\II_{1/2}[\bar\g_i,0]$ and $W^\II_{1/2}[\bar\g_i',0]$. When $\bar\g_i' \pp \bar\g_i$, they trivially preserve the same supercharges. Otherwise, when $\bar\g_i' \sp \bar\g_i$, they do not share any preserved supercharge.

  \item $W^\II_{1/2}[\bar\g_i,0]$ and $W^\II_{1/2}[0,\bar\g_\hi']$. They share 1/2 of the preserved supercharges
      \be \bar\g_i\bar\g'_\hj\th^{i\hj}_-, ~~ \ve_{ij}\ve_{\hk\hl}\g^i\g'^\hk\th^{j\hl}_+ \ee

  \item $W^\II_{1/2}[0,\bar\g_\hi]$ and $W^\II_{1/2}[0,\bar\g_\hi']$. When $\bar\g_\hi' \pp \bar\g_\hi$, they trivially preserve the same supercharges. Otherwise, they do not share any preserved supercharge.
\end{itemize}

These results are consistent with the examples in \cite{Lietti:2017gtc}.
We conclude that the only non--trivial pairs of different 1/2 BPS WLs preserving the same supercharges are
\bea
&& (W^\I_{1/2}[\bar\a_i,0], W^\II_{1/2}[\bar\g_i,0]) \quad {\textrm{with}} \quad \bar\a_i\g^i=0  \nonumber \\[0.3cm]
&& (W^\I_{1/2}[0,\bar\a_\hi], W^\II_{1/2}[0,\bar\g_\hi]) \quad {\textrm{with}} \quad  \bar\a_\hi\g^\hi=0
\eea

\section{Circular 1/4 BPS WLs in Euclidean $\mN=4$ SCSM theories}\label{appcir}

Linear WLs have constant expectation values, so they cannot be used to check non--trivially the matching between matrix model, field theory and holographic predictions. Non--trivial expectation values can be obtained for WLs on closed contours. However, since there are no spacelike BPS WLs in Minkowski spacetime \cite{Ouyang:2015ada}, we have to build--up circular BPS WLs in Euclidean space.

In this appendix, we generalize the procedure used for linear WLs in Minkowski spacetime to obtain the explicit form of euclidean circular 1/4 BPS WLs in $\mN=4$ circular quiver SCSM theories.

We use coordinates $x^\m=(x^1,x^2,x^3)$, metric $\d_{\m\n}=\diag(+++)$ and choose gamma matrices as
\be
\g^{\m~\b}_{~\a}=(-\s^2,\s^1,\s^3)
\ee
For the circle $x^\m=(\cos\t,\sin\t,0)$, we define the following Grassmann even spinors
\bea \label{g6}
&& u_{+\a} = \f{1}{\sqrt{2}} \left( \ba{cc} \ep^{-\f{\ii\t}2} \\ \ep^{\f{\ii\t}2} \ea \right), ~~
   u_{-\a} = \f{\ii}{\sqrt{2}} \left( \ba{cc} -\ep^{-\f{\ii\t}2} \\ \ep^{\f{\ii\t}2} \ea \right) \nn\\
&& u_{+}^{\a} = \f{1}{\sqrt{2}} \left( \ep^{\f{\ii\t}2}, -\ep^{-\f{\ii\t}2}  \right), ~~
   u_{-}^{\a} = \f{\ii}{\sqrt{2}} \left( \ep^{\f{\ii\t}2}, \ep^{-\f{\ii\t}2}  \right)
\eea

SUSY transformations and supercharges for $\mN=4$ SCSM theories in Euclidean space are formally the same as those in Minkowski spacetime (see eqs.  (\ref{appA1}), (\ref{e52})).

To be definite, we construct BPS WLs preserving supercharges
\be
\vth^{1\ho} = -\ii \g_3 \th^{1\ho}, ~~
\vth^{2\hw} = \ii \g_3 \th^{2\hw}
\ee
We find the bosonic 1/4 BPS WL $W_\bos$ with connection \cite{Ouyang:2015qma,Cooke:2015ila}
\bea \label{LbosE}
&& L_\bos = \diag ( \mA^{(1)}_\bos, \mB^{(2)}_\bos, \cdots, \mA^{(2r-1)}_\bos, \mB^{(2r)}_\bos )  \nn\\
&& \mA^{(2\ell-1)}_\bos =  A_\m^{(2\ell-1)} \dot x^\m
   + \f{2\pi\ii}{k} \Big( \phi_1^{(2\ell-2)}\bar\phi^1_{(2\ell-2)}
                      - \phi_2^{(2\ell-2)}\bar\phi^2_{(2\ell-2)}  \nn\\
&& \phantom{\mA^{(2\ell-1)}_\bos =}
                      + \phi_\ho^{(2\ell-1)}\bar\phi^\ho_{(2\ell-1)}
                      - \phi_\hw^{(2\ell-1)}\bar\phi^\hw_{(2\ell-1)}
                 \Big)|\dot x|  \\
&& \mB^{(2\ell)}_\bos =  B_\m^{(2\ell)}\dot x^\m
   + \f{2\pi\ii}{k} \Big(  \bar\phi^1_{(2\ell)}\phi_1^{(2\ell)}
                      - \bar\phi^2_{(2\ell)}\phi_2^{(2\ell)}
                      + \bar\phi^\ho_{(2\ell-1)}\phi_\ho^{(2\ell-1)}
                      - \bar\phi^\hw_{(2\ell-1)}\phi_\hw^{(2\ell-1)}
                 \Big)|\dot x|  \nn
\eea
For fermionic WLs we find superconnections $L_\fer$ of the form (\ref{l2}) or (\ref{l3}) with definitions
\bea
&& \mA^{(2\ell-1)} =  \mA^{(2\ell-1)}_\bos
   + \f{4\pi\ii}{k} \Big[
                      - \bar\a_\ho^{(2\ell-2)}\b^\ho_{(2\ell-2)} \phi_1^{(2\ell-2)}\bar\phi^1_{(2\ell-2)}
                      + \bar\g_\hw^{(2\ell-2)}\d^\hw_{(2\ell-2)} \phi_2^{(2\ell-2)}\bar\phi^2_{(2\ell-2)} \nn\\
&& \phantom{\mA^{(2\ell-1)} =}
                      - \bar\a_1^{(2\ell-1)}\b^1_{(2\ell-1)} \phi_\ho^{(2\ell-1)}\bar\phi^\ho_{(2\ell-1)}
                      + \bar\g_2^{(2\ell-1)}\d^2_{(2\ell-1)} \phi_\hw^{(2\ell-1)}\bar\phi^\hw_{(2\ell-1)}
                 \Big] |\dot x| \nn\\
&& \mB^{(2\ell)} =  \mB^{(2\ell)}_\bos
   + \f{4\pi\ii}{k} \Big[
                      - \bar\a_\ho^{(2\ell)}\b^\ho_{(2\ell)} \bar\phi^1_{(2\ell)}\phi_1^{(2\ell)}
                      + \bar\g_\hw^{(2\ell)}\d^\hw_{(2\ell)} \bar\phi^2_{(2\ell)}\phi_2^{(2\ell)} \nn\\
&& \phantom{\mB^{(2\ell)} =}
                      - \bar\a_1^{(2\ell-1)}\b^1_{(2\ell-1)} \bar\phi^\ho_{(2\ell-1)}\phi_\ho^{(2\ell-1)}
                      + \bar\g_2^{(2\ell-1)}\d^2_{(2\ell-1)} \bar\phi^\hw_{(2\ell-1)}\phi_\hw^{(2\ell-1)}
                 \Big] |\dot x| \nn\\
&& f_1^{(2\ell-1)} = \ii\sqrt{\f{4\pi}{k}} \Big( \bar\a_i^{(2\ell-1)} u_+
                                                - \bar\g_i^{(2\ell-1)} u_- \Big)\psi^i_{(2\ell-1)}|\dot x| \nn\\
&& f_1^{(2\ell)} = - \ii\sqrt{\f{4\pi}{k}} \bar \psi_{\hi}^{(2\ell)} \Big( u_-\b^\hi_{(2\ell)}
                                                                        + u_+ \d^\hi_{(2\ell)} \Big) |\dot x| \nn\\
&& f_2^{(2\ell-1)} =-  \ii\sqrt{\f{4\pi}{k}} \bar \psi_{i}^{(2\ell-1)} \Big( u_- \b^i_{(2\ell-1)}
                                                                          + u_+ \d^i_{(2\ell-1)} \Big)|\dot x| \nn\\
&& f_2^{(2\ell)} = \ii\sqrt{\f{4\pi}{k}} \Big( \bar\a_\hi^{(2\ell)} u_+
                                          - \bar\g_\hi^{(2\ell)} u_- \Big)\psi^\hi_{(2\ell)}|\dot x| \nn\\
&& h_1^{(2\ell-1)} = \f{4\pi\ii}{k} \Big(   \bar\a_1^{(2\ell-1)} \b^\ho_{(2\ell)} \phi_\ho^{(2\ell-1)}\bar\phi^1_{(2\ell)}
                                       - \bar\g_2^{(2\ell-1)} \d^\hw_{(2\ell)} \phi_\hw^{(2\ell-1)}\bar\phi^2_{(2\ell)} \Big)|\dot x| \nn\\
&& h_1^{(2\ell)} = \f{4\pi\ii}{k}  \Big(   \bar\a_1^{(2\ell+1)} \b^\ho_{(2\ell)} \bar\phi^1_{(2\ell)}\phi_\ho^{(2\ell+1)}
                                      - \bar\g_2^{(2\ell+1)} \d^\hw_{(2\ell)} \bar\phi^2_{(2\ell)}\phi_\hw^{(2\ell+1)} \Big) |\dot x|\nn\\
&& h_2^{(2\ell-1)} = \f{4\pi\ii}{k}  \Big(   \bar\a_\ho^{(2\ell)} \b^1_{(2\ell-1)} \phi_1^{(2\ell)} \bar\phi^\ho_{(2\ell-1)}
                                        - \bar\g_\hw^{(2\ell)} \d^2_{(2\ell-1)} \phi_2^{(2\ell)}\bar\phi^\hw_{(2\ell-1)} \Big) |\dot x|\nn\\
&& h_2^{(2\ell)} = \f{4\pi\ii}{k}  \Big(   \bar\a_\ho^{(2\ell)} \b^1_{(2\ell+1)} \bar\phi^\ho_{(2\ell+1)}\phi_1^{(2\ell)}
                                      - \bar\g_\hw^{(2\ell)} \d^2_{(2\ell+1)} \bar\phi^\hw_{(2\ell+1)}\phi_2^{(2\ell)} \Big)|\dot x|
\eea
They depend on $8r$ parameters
\be
\bar\a_1^{(2\ell-1)}, ~ \bar\a_\ho^{(2\ell)}, ~ \b^1_{(2\ell-1)}, ~ \b^\ho_{(2\ell)}, ~
\bar\g_2^{(2\ell-1)}, ~ \bar\g_\hw^{(2\ell)}, ~ \d^2_{(2\ell-1)}, ~ \d^\hw_{(2\ell)}, ~~
\ell=1,2,\cdots,r
\ee
subject to the following constraints
\bea
&& \bar\a_1^{(2\ell-1)} \d^2_{(2\ell-1)} =
   \bar\a_1^{(2\ell-1)} \d^\hw_{(2\ell-2)} =
   \bar\a_1^{(2\ell-1)} \d^\hw_{(2\ell)} = 0 \nn\\
&& \bar\a_\ho^{(2\ell)} \d^2_{(2\ell)} =
   \bar\a_\ho^{(2\ell)} \d^\hw_{(2\ell-1)} =
   \bar\a_\ho^{(2\ell)} \d^\hw_{(2\ell+1)} = 0 \nn\\
&& \bar\g_2^{(2\ell-1)} \b^1_{(2\ell-1)} =
   \bar\g_2^{(2\ell-1)} \b^\ho_{(2\ell-2)} =
   \bar\g_2^{(2\ell-1)} \b^\ho_{(2\ell)} = 0 \nn\\
&& \bar\g_\hw^{(2\ell)} \b^1_{(2\ell)} =
   \bar\g_\hw^{(2\ell)} \b^\ho_{(2\ell-1)} =
   \bar\g_\hw^{(2\ell)} \b^\ho_{(2\ell+1)} = 0
\eea
As in Minkowski case, solutions can be classified in terms of the number of free parameters they depend on. First, we find four classes of solutions depending on $4r$ free complex parameters
\bea
&& \textrm{Class I}: ~ \bar\g_2^{(2\ell-1)}=\bar\g_\hw^{(2\ell)}=\d^2_{(2\ell-1)}=\d^\hw_{(2\ell)} = 0 \nn\\
&& \textrm{Class II}: ~ \bar\a_1^{(2\ell-1)}=\bar\a_\ho^{(2\ell)}=\b^1_{(2\ell-1)}=\b^\ho_{(2\ell)} = 0 \nn\\
&& \textrm{Class III}: ~\b^1_{(2\ell-1)}=\b^\ho_{(2\ell)} =\d^2_{(2\ell-1)}=\d^\hw_{(2\ell)} = 0 \nn\\
&& \textrm{Class IV}: ~ \bar\a_1^{(2\ell-1)}=\bar\a_\ho^{(2\ell)} = \bar\g_2^{(2\ell-1)}=\bar\g_\hw^{(2\ell)}=0
\eea
In addition, we find extra solutions that do not fall into the previous four classes, being functions of a smaller number of parameters. All the considerations concerning the structure of  superconnessions being block or non--block--diagonal apply in this case exactly in the same manner as in Minkowski spacetime.

Also in the euclidean case fermionic 1/4 BPS WLs are cohomologically equivalent to the bosonic 1/4 BPS WLs. In fact, it is easy to prove that they satisfy the set of conditions (\ref{cohom}) with
\be
\mQ = -\z ( P^{1\ho} - \ii\g_3 S^{1\ho} ) - \z( P^{2\hw} + \ii\g_3S^{2\hw} ), ~~ \z^\a=(1,0), ~~ \k= 2 \ep^{-\ii\t}
\ee
and $\L$ of the form (\ref{lambda}) with
\bea
&& \l_1^{(2\ell-1)} =  \sr{\f{4\pi}{k}} \ep^{\f{\ii\t}2} \big( \ii \bar\a_1^{(2\ell-1)}\phi_\ho^{(2\ell-1)}
                                                                 + \bar\g_2^{(2\ell-1)}\phi_\hw^{(2\ell-1)} \big) \nn\\
&& \l_1^{(2\ell)} =  \sr{\f{4\pi}{k}} \ep^{\f{\ii\t}2} \big(  \b^\ho_{(2\ell)}\bar\phi^1_{(2\ell)}
                                                         -\ii \d^\hw_{(2\ell)}\bar\phi^2_{(2\ell)} \big) \nn\\
&& \l_2^{(2\ell-1)} =  \sr{\f{4\pi}{k}} \ep^{\f{\ii\t}2} \big( - \b^1_{(2\ell-1)}\bar\phi^\ho_{(2\ell-1)}
                                                            +\ii \d^2_{(2\ell-1)}\bar\phi^\hw_{(2\ell-1)} \big) \nn\\
&& \l_2^{(2\ell)} =  \sr{\f{4\pi}{k}} \ep^{\f{\ii\t}2} \big( - \ii \bar\a_\ho^{(2\ell)}\phi_1^{(2\ell)}
                                                                 - \bar\g_\hw^{(2\ell)}\phi_2^{(2\ell)} \big)
\eea

\providecommand{\href}[2]{#2}\begingroup\raggedright\endgroup


\providecommand{\href}[2]{#2}\begingroup\raggedright\begin{thebibliography}{10}

\bibitem{Gaiotto:2007qi}
D.~Gaiotto and X.~Yin, \emph{{Notes on superconformal Chern-Simons-Matter
  theories}},
  \href{http://dx.doi.org/10.1088/1126-6708/2007/08/056}{\emph{JHEP} {\bfseries
  0708} (2007) 056}, [\href{https://arxiv.org/abs/0704.3740}{{\ttfamily
  0704.3740}}].

\bibitem{Drukker:2009hy}
N.~Drukker and D.~Trancanelli, \emph{{A Supermatrix model for $\mathcal N=6$
  super Chern-Simons-matter theory}},
  \href{http://dx.doi.org/10.1007/JHEP02(2010)058}{\emph{JHEP} {\bfseries 1002}
  (2010) 058}, [\href{https://arxiv.org/abs/0912.3006}{{\ttfamily 0912.3006}}].

\bibitem{Pestun:2007rz}
V.~Pestun, \emph{{Localization of gauge theory on a four-sphere and
  supersymmetric Wilson loops}},
  \href{http://dx.doi.org/10.1007/s00220-012-1485-0}{\emph{Commun. Math. Phys.}
  {\bfseries 313} (2012) 71--129},
  [\href{https://arxiv.org/abs/0712.2824}{{\ttfamily 0712.2824}}].

\bibitem{Kapustin:2009kz}
A.~Kapustin, B.~Willett and I.~Yaakov, \emph{{Exact Results for Wilson Loops in
  Superconformal Chern-Simons Theories with Matter}},
  \href{http://dx.doi.org/10.1007/JHEP03(2010)089}{\emph{JHEP} {\bfseries 1003}
  (2010) 089}, [\href{https://arxiv.org/abs/0909.4559}{{\ttfamily 0909.4559}}].

\bibitem{Maldacena:1998im}
J.~M. Maldacena, \emph{{Wilson loops in large N field theories}},
  \href{http://dx.doi.org/10.1103/PhysRevLett.80.4859}{\emph{Phys. Rev. Lett.}
  {\bfseries 80} (1998) 4859--4862},
  [\href{https://arxiv.org/abs/hep-th/9803002}{{\ttfamily hep-th/9803002}}].

\bibitem{Rey:1998ik}
S.-J. Rey and J.-T. Yee, \emph{{Macroscopic strings as heavy quarks in large N
  gauge theory and anti-de Sitter supergravity}},
  \href{http://dx.doi.org/10.1007/s100520100799}{\emph{Eur. Phys. J.}
  {\bfseries C22} (2001) 379--394},
  [\href{https://arxiv.org/abs/hep-th/9803001}{{\ttfamily hep-th/9803001}}].

\bibitem{Ouyang:2015bmy}
H.~Ouyang, J.-B. Wu and J.-j. Zhang, \emph{{Construction and classification of
  novel BPS Wilson loops in quiver Chern-Simons-matter theories}},
  \href{http://dx.doi.org/10.1016/j.nuclphysb.2016.07.018}{\emph{Nucl. Phys.}
  {\bfseries B910} (2016) 496--527},
  [\href{https://arxiv.org/abs/1511.02967}{{\ttfamily 1511.02967}}].

\bibitem{Aharony:2008ug}
O.~Aharony, O.~Bergman, D.~L. Jafferis and J.~Maldacena, \emph{{N=6
  superconformal Chern-Simons-matter theories, M2-branes and their gravity
  duals}}, \href{http://dx.doi.org/10.1088/1126-6708/2008/10/091}{\emph{JHEP}
  {\bfseries 0810} (2008) 091},
  [\href{https://arxiv.org/abs/0806.1218}{{\ttfamily 0806.1218}}].

\bibitem{Hosomichi:2008jb}
K.~Hosomichi, K.-M. Lee, S.~Lee, S.~Lee and J.~Park, \emph{{N=5,6
  Superconformal Chern-Simons Theories and M2-branes on Orbifolds}},
  \href{http://dx.doi.org/10.1088/1126-6708/2008/09/002}{\emph{JHEP} {\bfseries
  0809} (2008) 002}, [\href{https://arxiv.org/abs/0806.4977}{{\ttfamily
  0806.4977}}].

\bibitem{Aharony:2008gk}
O.~Aharony, O.~Bergman and D.~L. Jafferis, \emph{{Fractional M2-branes}},
  \href{http://dx.doi.org/10.1088/1126-6708/2008/11/043}{\emph{JHEP} {\bfseries
  0811} (2008) 043}, [\href{https://arxiv.org/abs/0807.4924}{{\ttfamily
  0807.4924}}].

\bibitem{Ouyang:2015iza}
H.~Ouyang, J.-B. Wu and J.-j. Zhang, \emph{{Novel BPS Wilson loops in
  three-dimensional quiver Chern-Simons-matter theories}},
  \href{http://dx.doi.org/10.1016/j.physletb.2015.12.021}{\emph{Phys. Lett.}
  {\bfseries B753} (2016) 215--220},
  [\href{https://arxiv.org/abs/1510.05475}{{\ttfamily 1510.05475}}].

\bibitem{Drukker:2008zx}
N.~Drukker, J.~Plefka and D.~Young, \emph{{Wilson loops in 3-dimensional N=6
  supersymmetric Chern-Simons Theory and their string theory duals}},
  \href{http://dx.doi.org/10.1088/1126-6708/2008/11/019}{\emph{JHEP} {\bfseries
  0811} (2008) 019}, [\href{https://arxiv.org/abs/0809.2787}{{\ttfamily
  0809.2787}}].

\bibitem{Chen:2008bp}
B.~Chen and J.-B. Wu, \emph{{Supersymmetric Wilson Loops in N=6 Super
  Chern-Simons-matter theory}},
  \href{http://dx.doi.org/10.1016/j.nuclphysb.2009.09.015}{\emph{Nucl. Phys.}
  {\bfseries B825} (2010) 38--51},
  [\href{https://arxiv.org/abs/0809.2863}{{\ttfamily 0809.2863}}].

\bibitem{Rey:2008bh}
S.-J. Rey, T.~Suyama and S.~Yamaguchi, \emph{{Wilson Loops in Superconformal
  Chern-Simons Theory and Fundamental Strings in Anti-de Sitter Supergravity
  Dual}}, \href{http://dx.doi.org/10.1088/1126-6708/2009/03/127}{\emph{JHEP}
  {\bfseries 0903} (2009) 127},
  [\href{https://arxiv.org/abs/0809.3786}{{\ttfamily 0809.3786}}].

\bibitem{Lietti:2017gtc}
M.~Lietti, A.~Mauri, S.~Penati and J.-j. Zhang, \emph{{String theory duals of
  Wilson loops from Higgsing}},
  \href{http://dx.doi.org/10.1007/JHEP08(2017)030}{\emph{JHEP} {\bfseries 1708}
  (2017) 030}, [\href{https://arxiv.org/abs/1705.02322}{{\ttfamily
  1705.02322}}].

\bibitem{Griguolo:2012iq}
L.~Griguolo, D.~Marmiroli, G.~Martelloni and D.~Seminara, \emph{{The
  generalized cusp in ABJ(M) N = 6 Super Chern-Simons theories}},
  \href{http://dx.doi.org/10.1007/JHEP05(2013)113}{\emph{JHEP} {\bfseries 1305}
  (2013) 113}, [\href{https://arxiv.org/abs/1208.5766}{{\ttfamily 1208.5766}}].

\bibitem{Cardinali:2012ru}
V.~Cardinali, L.~Griguolo, G.~Martelloni and D.~Seminara, \emph{{New
  supersymmetric Wilson loops in ABJ(M) theories}},
  \href{http://dx.doi.org/10.1016/j.physletb.2012.10.051}{\emph{Phys. Lett.}
  {\bfseries B718} (2012) 615--619},
  [\href{https://arxiv.org/abs/1209.4032}{{\ttfamily 1209.4032}}].

\bibitem{Kim:2013oza}
N.~Kim, \emph{{Supersymmetric Wilson loops with general contours in ABJM
  theory}}, \href{http://dx.doi.org/10.1142/S0217732313501502}{\emph{Mod. Phys.
  Lett.} {\bfseries A28} (2013) 1350150},
  [\href{https://arxiv.org/abs/1304.7660}{{\ttfamily 1304.7660}}].

\bibitem{Bianchi:2014laa}
M.~S. Bianchi, L.~Griguolo, M.~Leoni, S.~Penati and D.~Seminara, \emph{{BPS
  Wilson loops and Bremsstrahlung function in ABJ(M): a two loop analysis}},
  \href{http://dx.doi.org/10.1007/JHEP06(2014)123}{\emph{JHEP} {\bfseries 1406}
  (2014) 123}, [\href{https://arxiv.org/abs/1402.4128}{{\ttfamily 1402.4128}}].

\bibitem{Correa:2014aga}
D.~H. Correa, J.~Aguilera-Damia and G.~A. Silva, \emph{{Strings in $AdS_4
  \times \mathbb{CP}^{3}$ Wilson loops in $\mathcal N=$6 super
  Chern-Simons-matter and bremsstrahlung functions}},
  \href{http://dx.doi.org/10.1007/JHEP06(2014)139}{\emph{JHEP} {\bfseries 1406}
  (2014) 139}, [\href{https://arxiv.org/abs/1405.1396}{{\ttfamily 1405.1396}}].

\bibitem{Gaiotto:2008sd}
D.~Gaiotto and E.~Witten, \emph{{Janus Configurations, Chern-Simons Couplings,
  And The theta-Angle in N=4 Super Yang-Mills Theory}},
  \href{http://dx.doi.org/10.1007/JHEP06(2010)097}{\emph{JHEP} {\bfseries 1006}
  (2010) 097}, [\href{https://arxiv.org/abs/0804.2907}{{\ttfamily 0804.2907}}].

\bibitem{Hosomichi:2008jd}
K.~Hosomichi, K.-M. Lee, S.~Lee, S.~Lee and J.~Park, \emph{{N=4 Superconformal
  Chern-Simons Theories with Hyper and Twisted Hyper Multiplets}},
  \href{http://dx.doi.org/10.1088/1126-6708/2008/07/091}{\emph{JHEP} {\bfseries
  0807} (2008) 091}, [\href{https://arxiv.org/abs/0805.3662}{{\ttfamily
  0805.3662}}].

\bibitem{Benna:2008zy}
M.~Benna, I.~Klebanov, T.~Klose and M.~Smedback, \emph{{Superconformal
  Chern-Simons Theories and AdS$_4$/CFT$_3$ Correspondence}},
  \href{http://dx.doi.org/10.1088/1126-6708/2008/09/072}{\emph{JHEP} {\bfseries
  0809} (2008) 072}, [\href{https://arxiv.org/abs/0806.1519}{{\ttfamily
  0806.1519}}].

\bibitem{Imamura:2008nn}
Y.~Imamura and K.~Kimura, \emph{{On the moduli space of elliptic
  Maxwell-Chern-Simons theories}},
  \href{http://dx.doi.org/10.1143/PTP.120.509}{\emph{Prog. Theor. Phys.}
  {\bfseries 120} (2008) 509--523},
  [\href{https://arxiv.org/abs/0806.3727}{{\ttfamily 0806.3727}}].

\bibitem{Terashima:2008ba}
S.~Terashima and F.~Yagi, \emph{{Orbifolding the Membrane Action}},
  \href{http://dx.doi.org/10.1088/1126-6708/2008/12/041}{\emph{JHEP} {\bfseries
  0812} (2008) 041}, [\href{https://arxiv.org/abs/0807.0368}{{\ttfamily
  0807.0368}}].

\bibitem{Imamura:2008dt}
Y.~Imamura and K.~Kimura, \emph{{N=4 Chern-Simons theories with auxiliary
  vector multiplets}},
  \href{http://dx.doi.org/10.1088/1126-6708/2008/10/040}{\emph{JHEP} {\bfseries
  0810} (2008) 040}, [\href{https://arxiv.org/abs/0807.2144}{{\ttfamily
  0807.2144}}].

\bibitem{Cooke:2015ila}
M.~Cooke, N.~Drukker and D.~Trancanelli, \emph{{A profusion of $1/2$ BPS Wilson
  loops in $\mathcal{N}=4$ Chern-Simons-matter theories}},
  \href{http://dx.doi.org/10.1007/JHEP10(2015)140}{\emph{JHEP} {\bfseries 1510}
  (2015) 140}, [\href{https://arxiv.org/abs/1506.07614}{{\ttfamily
  1506.07614}}].

\bibitem{Ouyang:2015qma}
H.~Ouyang, J.-B. Wu and J.-j. Zhang, \emph{{Supersymmetric Wilson loops in
  $\mathcal N=4$ super Chern-Simons-matter theory}},
  \href{http://dx.doi.org/10.1007/JHEP11(2015)213}{\emph{JHEP} {\bfseries 1511}
  (2015) 213}, [\href{https://arxiv.org/abs/1506.06192}{{\ttfamily
  1506.06192}}].

\bibitem{Gaiotto:2008cg}
D.~Gaiotto, S.~Giombi and X.~Yin, \emph{{Spin Chains in N=6 Superconformal
  Chern-Simons-Matter Theory}},
  \href{http://dx.doi.org/10.1088/1126-6708/2009/04/066}{\emph{JHEP} {\bfseries
  0904} (2009) 066}, [\href{https://arxiv.org/abs/0806.4589}{{\ttfamily
  0806.4589}}].

\bibitem{Terashima:2008sy}
S.~Terashima, \emph{{On M5-branes in N=6 Membrane Action}},
  \href{http://dx.doi.org/10.1088/1126-6708/2008/08/080}{\emph{JHEP} {\bfseries
  0808} (2008) 080}, [\href{https://arxiv.org/abs/0807.0197}{{\ttfamily
  0807.0197}}].

\bibitem{Bandres:2008ry}
M.~A. Bandres, A.~E. Lipstein and J.~H. Schwarz, \emph{{Studies of the ABJM
  Theory in a Formulation with Manifest SU(4) R-Symmetry}},
  \href{http://dx.doi.org/10.1088/1126-6708/2008/09/027}{\emph{JHEP} {\bfseries
  0809} (2008) 027}, [\href{https://arxiv.org/abs/0807.0880}{{\ttfamily
  0807.0880}}].

\bibitem{Lee:2010hk}
K.-M. Lee and S.~Lee, \emph{{1/2-BPS Wilson loops and vortices in ABJM model}},
  \href{http://dx.doi.org/10.1007/JHEP09(2010)004}{\emph{JHEP} {\bfseries 1009}
  (2010) 004}, [\href{https://arxiv.org/abs/1006.5589}{{\ttfamily 1006.5589}}].

\bibitem{Marino:2009jd}
M.~Marino and P.~Putrov, \emph{{Exact Results in ABJM Theory from Topological
  Strings}}, \href{http://dx.doi.org/10.1007/JHEP06(2010)011}{\emph{JHEP}
  {\bfseries 1006} (2010) 011},
  [\href{https://arxiv.org/abs/0912.3074}{{\ttfamily 0912.3074}}].

\bibitem{Drukker:2010nc}
N.~Drukker, M.~Marino and P.~Putrov, \emph{{From weak to strong coupling in
  ABJM theory}},
  \href{http://dx.doi.org/10.1007/s00220-011-1253-6}{\emph{Commun. Math. Phys.}
  {\bfseries 306} (2011) 511--563},
  [\href{https://arxiv.org/abs/1007.3837}{{\ttfamily 1007.3837}}].

\bibitem{Ouyang:2015hta}
H.~Ouyang, J.-B. Wu and J.-j. Zhang, \emph{{Exact results for Wilson loops in
  orbifold ABJM theory}},
  \href{http://dx.doi.org/10.1088/1674-1137/40/8/083101}{\emph{Chin. Phys.}
  {\bfseries C40} (2016) 083101},
  [\href{https://arxiv.org/abs/1507.00442}{{\ttfamily 1507.00442}}].

\bibitem{Griguolo:2015swa}
L.~Griguolo, M.~Leoni, A.~Mauri, S.~Penati and D.~Seminara, \emph{{Probing
  Wilson loops in ${\mathcal N}=4$ Chern-Simons-matter theories at weak
  coupling}},
  \href{http://dx.doi.org/10.1016/j.physletb.2015.12.018}{\emph{Phys. Lett.}
  {\bfseries B753} (2016) 500--505},
  [\href{https://arxiv.org/abs/1510.08438}{{\ttfamily 1510.08438}}].

\bibitem{Bianchi:2016vvm}
M.~S. Bianchi, L.~Griguolo, M.~Leoni, A.~Mauri, S.~Penati and D.~Seminara,
  \emph{{The quantum 1/2 BPS Wilson loop in ${\cal N}=4$ Chern-Simons-matter
  theories}}, \href{http://dx.doi.org/10.1007/JHEP09(2016)009}{\emph{JHEP}
  {\bfseries 1609} (2016) 009},
  [\href{https://arxiv.org/abs/1606.07058}{{\ttfamily 1606.07058}}].

\bibitem{Bianchi:2016yzj}
M.~S. Bianchi, L.~Griguolo, M.~Leoni, A.~Mauri, S.~Penati and D.~Seminara,
  \emph{{Framing and localization in Chern-Simons theories with matter}},
  \href{http://dx.doi.org/10.1007/JHEP06(2016)133}{\emph{JHEP} {\bfseries 1606}
  (2016) 133}, [\href{https://arxiv.org/abs/1604.00383}{{\ttfamily
  1604.00383}}].

\bibitem{Bianchi:2016gpg}
M.~S. Bianchi, \emph{{A note on multiply wound BPS Wilson loops in ABJM}},
  \href{http://dx.doi.org/10.1007/JHEP09(2016)047}{\emph{JHEP} {\bfseries 1609}
  (2016) 047}, [\href{https://arxiv.org/abs/1605.01025}{{\ttfamily
  1605.01025}}].

\bibitem{Bianchi:2013rma}
M.~S. Bianchi, G.~Giribet, M.~Leoni and S.~Penati, \emph{{The 1/2 BPS Wilson
  loop in ABJ(M) at two loops: The details}},
  \href{http://dx.doi.org/10.1007/JHEP10(2013)085}{\emph{JHEP} {\bfseries 1310}
  (2013) 085}, [\href{https://arxiv.org/abs/1307.0786}{{\ttfamily 1307.0786}}].

\bibitem{Griguolo:2013sma}
L.~Griguolo, G.~Martelloni, M.~Poggi and D.~Seminara, \emph{{Perturbative
  evaluation of circular 1/2 BPS Wilson loops in N = 6 Super Chern-Simons
  theories}}, \href{http://dx.doi.org/10.1007/JHEP09(2013)157}{\emph{JHEP}
  {\bfseries 1309} (2013) 157},
  [\href{https://arxiv.org/abs/1307.0787}{{\ttfamily 1307.0787}}].

\bibitem{work}
A.~Mauri, S.~Penati and J.-j. Zhang.

\bibitem{Lewkowycz:2013laa}
A.~Lewkowycz and J.~Maldacena, \emph{{Exact results for the entanglement
  entropy and the energy radiated by a quark}},
  \href{http://dx.doi.org/10.1007/JHEP05(2014)025}{\emph{JHEP} {\bfseries 1405}
  (2014) 025}, [\href{https://arxiv.org/abs/1312.5682}{{\ttfamily 1312.5682}}].

\bibitem{Correa:2012hh}
D.~Correa, J.~Maldacena and A.~Sever, \emph{{The quark anti-quark potential and
  the cusp anomalous dimension from a TBA equation}},
  \href{http://dx.doi.org/10.1007/JHEP08(2012)134}{\emph{JHEP} {\bfseries 1208}
  (2012) 134}, [\href{https://arxiv.org/abs/1203.1913}{{\ttfamily 1203.1913}}].

\bibitem{Drukker:2012de}
N.~Drukker, \emph{{Integrable Wilson loops}},
  \href{http://dx.doi.org/10.1007/JHEP10(2013)135}{\emph{JHEP} {\bfseries 1310}
  (2013) 135}, [\href{https://arxiv.org/abs/1203.1617}{{\ttfamily 1203.1617}}].

\bibitem{Gromov:2014eha}
N.~Gromov and G.~Sizov, \emph{{Exact Slope and Interpolating Functions in N=6
  Supersymmetric Chern-Simons Theory}},
  \href{http://dx.doi.org/10.1103/PhysRevLett.113.121601}{\emph{Phys. Rev.
  Lett.} {\bfseries 113} (2014) 121601},
  [\href{https://arxiv.org/abs/1403.1894}{{\ttfamily 1403.1894}}].

\bibitem{Cavaglia:2014exa}
A.~Cavagli\`a, D.~Fioravanti, N.~Gromov and R.~Tateo, \emph{{Quantum Spectral
  Curve of the $\mathcal N=$ 6 Supersymmetric Chern-Simons Theory}},
  \href{http://dx.doi.org/10.1103/PhysRevLett.113.021601}{\emph{Phys. Rev.
  Lett.} {\bfseries 113} (2014) 021601},
  [\href{https://arxiv.org/abs/1403.1859}{{\ttfamily 1403.1859}}].

\bibitem{Bombardelli:2017vhk}
D.~Bombardelli, A.~Cavagli\'a, D.~Fioravanti, N.~Gromov and R.~Tateo,
  \emph{{The full Quantum Spectral Curve for $AdS_4/CFT_3$}},
  \href{http://dx.doi.org/10.1007/JHEP09(2017)140}{\emph{JHEP} {\bfseries 1709}
  (2017) 140}, [\href{https://arxiv.org/abs/1701.00473}{{\ttfamily
  1701.00473}}].

\bibitem{Bianchi:2017svd}
M.~S. Bianchi, L.~Griguolo, A.~Mauri, S.~Penati, M.~Preti and D.~Seminara,
  \emph{{Towards the exact Bremsstrahlung function of ABJM theory}},
  \href{http://dx.doi.org/10.1007/JHEP08(2017)022}{\emph{JHEP} {\bfseries 1708}
  (2017) 022}, [\href{https://arxiv.org/abs/1705.10780}{{\ttfamily
  1705.10780}}].

\bibitem{Ouyang:2015ada}
H.~Ouyang, J.-B. Wu and J.-j. Zhang, \emph{{BPS Wilson loops in Minkowski
  spacetime and Euclidean space}},
  \href{http://dx.doi.org/10.1140/epjc/s10052-015-3834-6}{\emph{Eur. Phys. J.}
  {\bfseries C75} (2015) 606},
  [\href{https://arxiv.org/abs/1504.06929}{{\ttfamily 1504.06929}}].

\end{thebibliography}\endgroup

\end{document}